\documentclass{pasj01}

\Received{}
\Accepted{2021/05/10}
 
\usepackage{rotating}
 
\begin{document} 

\title{Three-dimensional dust geometry of the LMC H\,\emissiontype{I} ridge region as revealed by the IRSF/SIRIUS survey}

\author{Takuya \textsc{Furuta}\altaffilmark{1}$^{*}$}%
\altaffiltext{1}{Graduate School of Science, Nagoya University, Furo-cho, Chikusa-ku, Nagoya, Aichi 464-8602, Japan}
\email{t.furuta@u.phys.nagoya-u.ac.jp}

\author{Hidehiro \textsc{Kaneda}\altaffilmark{1}}%

\author{Takuma \textsc{Kokusho}\altaffilmark{1}}%


\author{Yasushi \textsc{Nakajima}\altaffilmark{2}}%
\altaffiltext{2}{National Astronomical Observatory of Japan, 2-21-1 Osawa, Mitaka, Tokyo 181-8588, Japan}

\author{Yasuo \textsc{Fukui},\altaffilmark{1}}

\author{Kisetsu \textsc{Tsuge},\altaffilmark{1}}


\maketitle

\begin{abstract}
We present a new method to evaluate the dust extinction ($A_V$) along the line of sight using the InfraRed Survey Facility (IRSF) near-infrared (NIR) data of the Large Magellanic Cloud (LMC) H\,\emissiontype{I} ridge region.
In our method, we estimate an $A_V$ value for each star from the NIR color excess and sort them from bluer to redder in each line of sight.
Using the percentile values of the sorted $A_V$, we newly construct the three-dimensional $A_V$ map.
We compare the resultant $A_V$ map with the total hydrogen column density $N$(H) traced by velocity-resolved H\,\emissiontype{I} and CO observations. 
In the LMC H\,\emissiontype{I} ridge region, Fukui et al. (2017, PASJ, 69, L5) find two velocity components and an intermediate velocity one bridging them.
Comparing our three-dimensional $A_V$ maps with $N$(H) maps at the different velocities, we find that the dust geometry is consistent with the scenario of the on-going gas collision between the two velocities as suggested in the previous study.
In addition, we find difference by a factor of 2 in $A_V$/$N$(H) between the two velocity components, which suggests that inflow gas from the Small Magellanic Clouds (SMC) is mixed in this region.
As a whole, our results support the triggered star formation in 30 Doradus due to the large-scale gas collision caused by tidal interaction between the LMC and the SMC.
\end{abstract}

\section{Introduction}
The Large Magellanic Cloud (LMC) represents an ideal laboratory for studying the interstellar medium (ISM) among external galaxies.
Because of its proximity ($\simeq$ 50 $\rm kpc$; \cite{proximity}) and its nearly face-on orientation ($i\sim35^{\circ}$; \cite{inclination}), we can perform spatially well-resolved studies of the ISM in almost two dimensions.
In addition, the LMC is a suitable site for studying massive star formation in low metallicity environments ($Z \sim$ 0.5 $\rm Z_{\odot}$; \cite{metal}).
In the LMC H\,\emissiontype{I} ridge region, several massive star formations have been reported in 30 Doradus (30 Dor; e.g., \cite{fukui_2017}; \cite{30dor_alma}) and N159 (e.g., \cite{n159_fukui}; \cite{n159_tokuda}).
In particular, 30 Dor is the most extreme starburst in the Local Universe. 
30 Dor harbors about 400 O-type/Wolf-Rayet stars, primarily powered by the central super star cluster R136 (\cite{r136_star}).
R136 has the mass of $5.0 \times 10^4\ \rm M_{\odot}$, 10 times larger than that of a super star cluster in our Galaxy and hosts the most massive stars heavier than $200\ \rm M_{\odot}$ (\cite{30dor_star}; \cite{crowther_2016}; \cite{schneider}). 

A recent study suggests that massive O- and B-type stars in R136 were formed by collision of H\,\emissiontype{I} clouds (\cite{fukui_2017}).
In their study, two velocity components of H\,\emissiontype{I} clouds and an intermediate velocity one bridging them are identified, from which they suggest that massive star formation in 30 Dor has been triggered by collision between these two velocity components. 
In addition, they compare the dust optical depth at 353 GHz, $\tau_{353}$, with the H\,\emissiontype{I} intensity $W$(H\,\emissiontype{I}), and find that $\tau_{353}$/$W$(H\,\emissiontype{I}) in the H\,\emissiontype{I} ridge region is lower than that outside the H\,\emissiontype{I} ridge region.
They suggest that the lower $\tau_{353}$/$W$(H\,\emissiontype{I}) ratio is caused by the contamination of inflow gas from the Small Magellanic Cloud (SMC), which is known to have even lower metallicity gas than the LMC, based on the result of the numerical simulation (\cite{bekkia}; \cite{shock_sim}; \cite{yozin}). 
Hence the massive star formation in the H\,\emissiontype{I} ridge region is likely to have been triggered by the galactic interaction, and thus studying the dust/gas ratios of the H\,\emissiontype{I} ridge region is important to identify such inflow gases from the SMC. 
One of the most reliable indicators of the dust column density is the near-infrared (NIR) dust extinction because it does not depend much on the dust temperature which is difficult to estimate since irrelevant components can be contaminated along the line of sight.

The method to evaluate the NIR dust extinction was originally introduced by \citet{nice}.  
In this method, the $H-K$ color excess is estimated from the mean $H-K$ color in a reference field (i.e., extinction-free region), which is so called Near-Infrared Color Excess (NICE) method. 
\citet{nicer} developed the NICE method by combining the $H-K$ and $J-H$ colors, which is called the NICER (NICE revised) method. 
Furthermore, \citet{dobashi} proposed the ``$X$ percentile metho''.
In this method, first, the colors are sorted from blue to red.
Then the color excess is estimated from the difference between the $X$ percentile of the color ($X=0$--$100\%$. $X=100\%$ denotes the reddest star) and that of the reference field.
The $X$ percentile method has the advantage that it is less affected by the contamination of the foreground stars causing under-estimation of the dust extinction. 
By using these methods, NIR dust extinction maps in the LMC were constructed based on the 2MASS data (\cite{extinction_nice} using the NICER method; \cite{dobashi} using the $X$ percentile method). 

\citet{furuta} derived the NIR dust extinction map of the LMC HI ridge region from the difference between the mean observed color and intrinsic colors accounting for several stellar spectral types.
They compared the spatial distribution of the dust extinction with that of the hydrogen column density of different velocity components, and revealed differences in the dust/gas ratios between multiple velocity components, indicating that inflow gas from the SMC is present in the LMC H\,\emissiontype{I} ridge region. 
In addition, three-dimensional dust geometry was suggested based on the spatial (anti-)correlation between the dust extinction and hydrogen column density, which supported the gas collisions around 30 Dor.
In this way, \citet{furuta} demonstrated that the NIR extinction is a powerful tool to evaluate not only the dust/gas ratio but also dust geometry.
However, the dust extinction estimated by \citet{furuta} should be more or less under-estimated unless all the stars are located behind the clouds, because the color excess is calculated from the mean observed colors.

In this paper, we develop the method used in \citet{furuta} by adopting the $X$ percentile method to evaluate the dust extinction more precisely along the line of sight and to reveal the three-dimensional dust geometry.
Using the resultant dust extinction map and gas maps of different velocity components, we discuss the dust distributions and dust/gas ratios of the multiple cloud components with different velocities. 

\section{The data}
\subsection{IRSF/SIRIUS point source catalog}
We used the NIR ($J$, $H$ and $K_{S}$ bands) photometric catalog for the Magellanic Clouds (hereafter the IRSF catalog) obtained with the SIRIUS camera on the InfraRed Survey Facility (IRSF) 1.4 m telescope (\cite{irsf}). 
In the IRSF catalog, \citet{furuta} found systematic photometric errors due to the reflection of the incident twilight sky illumination inside the SIRIUS camera, and therefore reprocessed the raw data of the IRSF catalog by using the corrected flat-field images (for details of the analysis, see \cite{furuta}).
In the present study, we selected data from this newly calibrated IRSF catalog according to the following criteria: (1) the magnitudes of the $J$, $H$ and $K_{S}$ bands are brighter than 18.8, 17.8 and 16.6 mag, respectively, corresponding to the $10\sigma$ limiting magnitudes of the original IRSF catalog, (2) ``quality flag'' in the original IRSF catalog is 1 in all the bands, indicating that the shapes of sources are ``point like'' and (3) the number of combined dithered images is larger than 8 in all the bands. 

\subsection{Sample selection}
\begin{figure}
 \begin{center}
  \includegraphics[width=8cm]{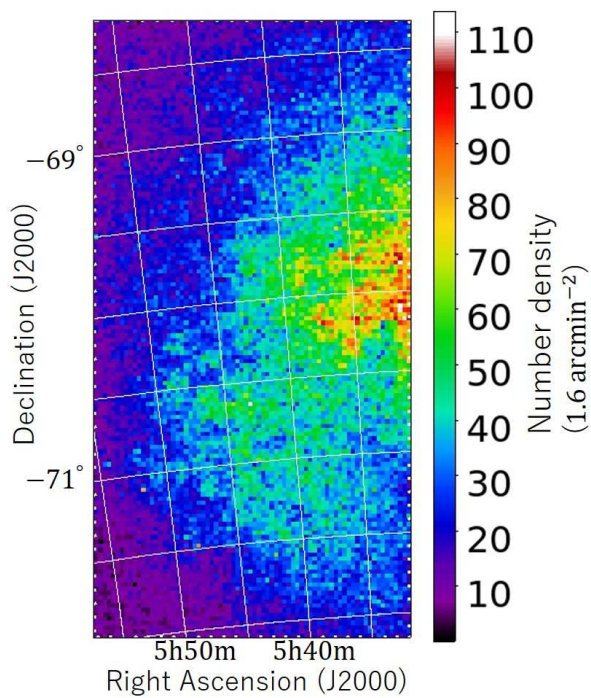} 
 \end{center}
\caption{Number density map of the selected sources of the LMC H\,\emissiontype{I} ridge region. The number of the stars included in each pixel of \timeform{1'.6}$\times$\timeform{1'.6} is shown in the color scale.}\label{fig:number_density}
\end{figure}

The IRSF catalog is contaminated with Galactic foreground stars, which causes under-estimation of the dust extinction.
Thus, we identify stellar populations using a color-magnitude diagram (CM diagram) according to the method used in \citet{furuta}.
The authors classified stellar populations into three categories which mainly consist of main sequence stars, red giant branch (RGB) stars and Galactic foreground stars.
The boundary lines between the three categories are determined on the CM diagram ($J-K$ vs. $K$) in their study.
By using the same boundary lines, we classify the stellar populations, and use only the objects classified as RGB stars from the CM diagram.
Figure \ref{fig:number_density} shows the number density map of the selected RGB stars in the H\,\emissiontype{I} ridge region.

\section{Method}\label{sec:method}
We describe our new method to measure the dust extinction along the line of sight in the LMC.
The key is to take percentile values of the $A_V$ distribution in each line of sight based on the $X$ percentile method.
In the following subsections, we explain our procedure step by step.
\begin{figure}
 \begin{center}
  \includegraphics[width=8cm]{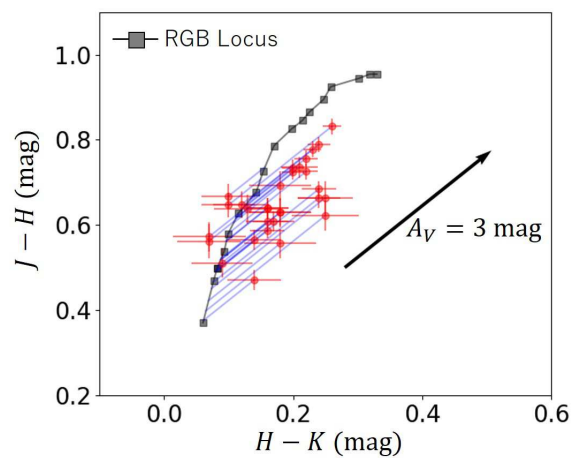} 
 \end{center}
\caption{Example of the color-color ($H-K$ vs. $J-H$) diagrams of the sources selected within a spatial bin of \timeform{1'.6}$\times$\timeform{1'.6}. Red circles and black squares are observed and intrinsic colors of RGB stars, respectively. The black arrow shows the reddening vector corresponding to $A_{V}=3 \rm \ mag$. Blue solid lines show the separation length between the observed and intrinsic colors of each star along the reddening vector.}\label{fig:cc_example}
\end{figure}
\subsection{Measuring $A_V$ for each star}\label{sec:measure_av}
We measure the dust extinction ($A_V$) for each star according to the method used in \citet{furuta}.
In this method, $A_V$ is calculated from the difference between the observed and intrinsic colors on a color-color diagram (CC diagram).
In figure \ref{fig:cc_example}, we present a CC diagram ($H-K$ vs. $J-H$). 
The line connected with the black squares is the locus of the intrinsic colors of RGB stars (\cite{tr_giant}).
The black arrow in figure \ref{fig:cc_example} is the reddening vector obtained from \citet{red_law2}, which is determined by the NIR photometry in the Johnson system.
To unify the photometric system, the observed and intrinsic colors are converted into the Johnson system according to the conversion equations suggested by \citet{color_corr} and \citet{color_corr2}.
From the reddening law of \citet{red_law2}, we have
\begin{equation}
A_{V}=10.9E(J - H)
\label{eq:excess1}
\end{equation}
and
 \begin{equation}
A_{V}=13.2E(H - K),
\label{eq:excess2}
\end{equation}
where $E(J - H)$ and $E(H - K)$ are the color excess of the $J-H$ and $H - K$ colors, respectively.
Using equations (\ref{eq:excess1}) and (\ref{eq:excess2}), we estimate $A_V$ for each star from the separation length between the intrinsic and observed colors of each RGB star along the reddening vector on the CC diagram.

Some of the observed colors in figure \ref{fig:cc_example} are bluer than the intrinsic colors.
In this case, we assign negative extinction corresponding to the separation length between the observed and intrinsic colors. 
In case that no intersection point exists between the reddening vector and the locus of the intrinsic colors, ``not a number'' is assigned to the corresponding star.

 \subsection{Calculation of $A_V$ distribution}\label{sec:percentile}
\begin{figure*}
 \begin{center}
  \includegraphics[width=16cm]{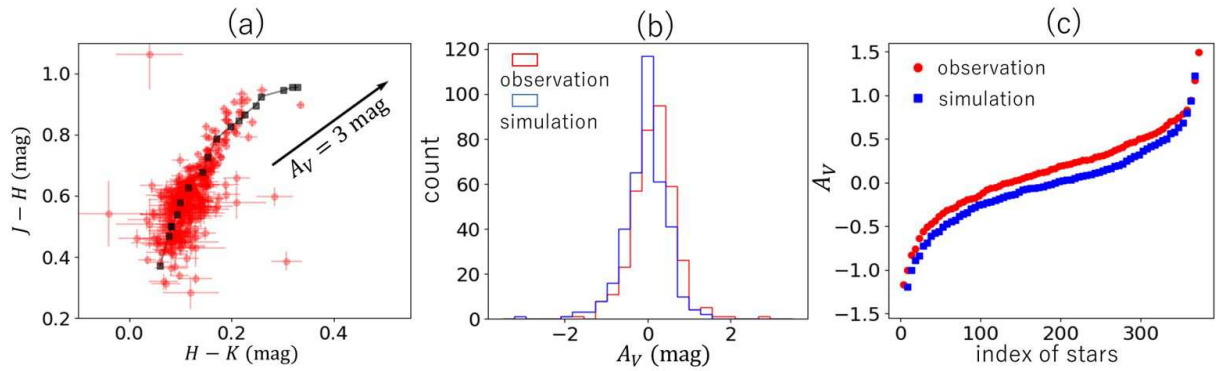} 
 \end{center}
\caption{(a) Color-color diagram ($H - K$ vs. $J - H$) of stars within a spatial bin of \timeform{5'} at $(\alpha, \delta)_{\rm J2000.0} =$ (\timeform{5h32m}, $-$\timeform{69D16'}). (b) Histogram of observed (red) and simulated (blue) $A_{V}$ values in the same region as in panel (a). (c) $A_V$ distribution as a function of index of stars obtained from the same data as in panel (a). Red circles and blue squares are observed and simulated $A_V$ plotted at every 5 stars, respectively.}\label{fig:eg_percentile}
\end{figure*}

 For each spatial bin, we sort the stars with respect to their $A_{V}$ values estimated in section \ref{sec:measure_av}.
 In figure \ref{fig:eg_percentile}a, we show an example of the CC diagram of the stars included in a spatial bin of \timeform{5'}$\times$\timeform{5'} at $(\alpha, \delta)_{\rm J2000.0} =$ (\timeform{5h32m}, $-$\timeform{69D16'}).
 This spatial bin (the position is plotted in figures \ref{fig:av_w_gas} and \ref{fig:tau_w_av}) is located in the LMC stellar bar region and has a low hydrogen column density ($<5\times10^{20}\  \rm{cm^{-2}}$).
 We select this region as an example for demonstrating the calculation of $A_V$ distribution where we expect  that no appreciable dust cloud is located along the line of sight.
 
 The estimated $A_V$ distribution in the bin is shown by the red histogram in figure \ref{fig:eg_percentile}b.
 To check the $A_V$ distribution quantitatively, we plot individual $A_V$ at every 5 stars as shown by the red circles in figure \ref{fig:eg_percentile}c.
The slope of the $A_V$ distribution in figure \ref{fig:eg_percentile}c indicates the presence of dust along the line of sight.
 However, even when there is no dust cloud along the line of sight, the slope is not zero due to the $A_V$ scatter caused by photometric errors.
 To evaluate this effect, we simulate the $A_V$ scatter based on the photometric errors as shown by the blue histogram and blue squares in figures \ref{fig:eg_percentile}b and \ref{fig:eg_percentile}c, respectively.
 As can be seen in the figures, the simulated $A_V$ is scattered around zero due to the photometric errors, the distribution of which is quite similar to the observed $A_V$ distribution.
 Therefore, we eliminate this effect from the observed data to evaluate the intrinsic dust distributions along the line of sight.
 In the following subsection, we describe the method to simulate the $A_V$ scatter and to eliminate this effect.
\subsection{$A_V$ scatter caused by photometric errors}\label{sec:sim_av}
\subsubsection{Calibration of the measured photometric errors}\label{sec:corr_error}
\begin{figure*}
\begin{center}
 \includegraphics[width=16cm]{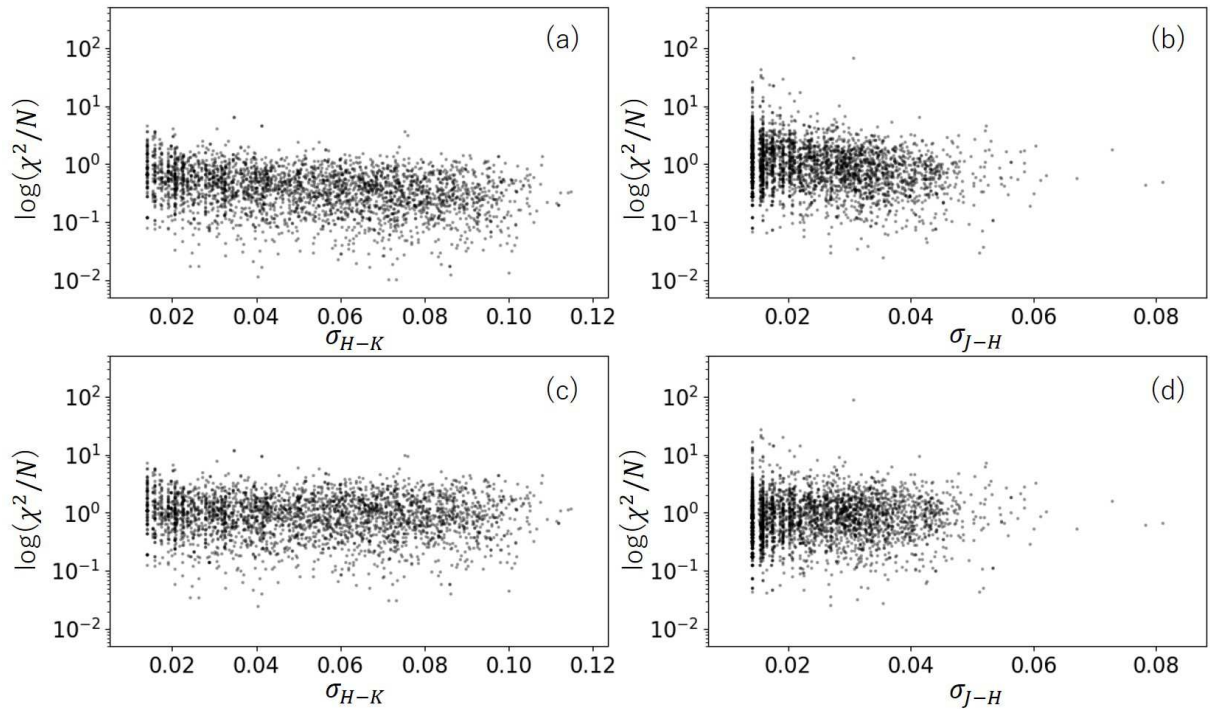} 
\end{center}
\caption{Logarithmic plots of $\chi^2/N$ (defined by equation \ref{eq:calib_err} in the text) vs. photometric errors of (a) the $H-K$ color and (b) the $J-H$ color for the stars in the present data which are observed more than four times. Panels (c) and (d) are the same diagrams as in panels (a) and (b), respectively, but those after re-scaling their photometric errors by applying the scaling factors $F$}\label{fig:err_calib}
\end{figure*}

If the photometric errors in the IRSF catalog are over- or under-estimated for some reason, we cannot appropriately evaluate the $A_V$ scatter caused by photometric errors. 
When photometric errors are appropriate values, the logarithm of $\chi^2/N$ should be close to zero for stars observed $N$ times (e.g., \cite{variable_1}; \cite{variable_2}). 
Here, $\chi^2$ is defined as
\begin{equation}
\chi^2=\sum_{}^N \left(\frac{m-\bar{m}}{\sigma_{\rm m}}\right)^2,
\label{eq:calib_err}
\end{equation}
where $\bar{m}$ is the weighted mean value of the colors of $m$ observed $N$ times for a star ($m=H-K$ or $J-H$), and $\sigma_{m}$ is the photometric error.
We calculate $\sigma_{H-K}=\sqrt{{\sigma_{H}}^2+{\sigma_{K}}^2}$ and $\sigma_{J-H}=\sqrt{{\sigma_{J}}^2+{\sigma_{H}}^2}$, where $\sigma_{J}$, $\sigma_{H}$ and $\sigma_{K}$ are the photometric errors of the $J$, $H$ and $K$ bands, respectively. 
By using the stars which are observed more than four times (i.e., $N>4$ in equation \ref{eq:calib_err}), we plot $\chi^2/N$ of the $H-K$ and $J-H$ colors as a function of their errors of $\sigma_{H-K}$ and $\sigma_{J-H}$ in figures \ref{fig:err_calib}a and \ref{fig:err_calib}b, respectively.
From figures \ref{fig:err_calib}a and \ref{fig:err_calib}b, we see that  $\chi^2/N$ becomes lower for higher $\sigma_{H-K}$ and $\sigma_{J-H}$, showing that photometric errors are over-estimated for stars which have large photometric errors.
Likely causes of the over-estimation are imperfect Point Spread Function (PSF) photometry, and/or over-estimation of the zero-point uncertainty.
To solve this problem, we calculate the scaling factor $F$ which is applied to $\sigma_{H-K}$ and $\sigma_{J-H}$, following the method of \citet{variable_1}. 

\begin{figure*}
\begin{center}
 \includegraphics[width=16cm]{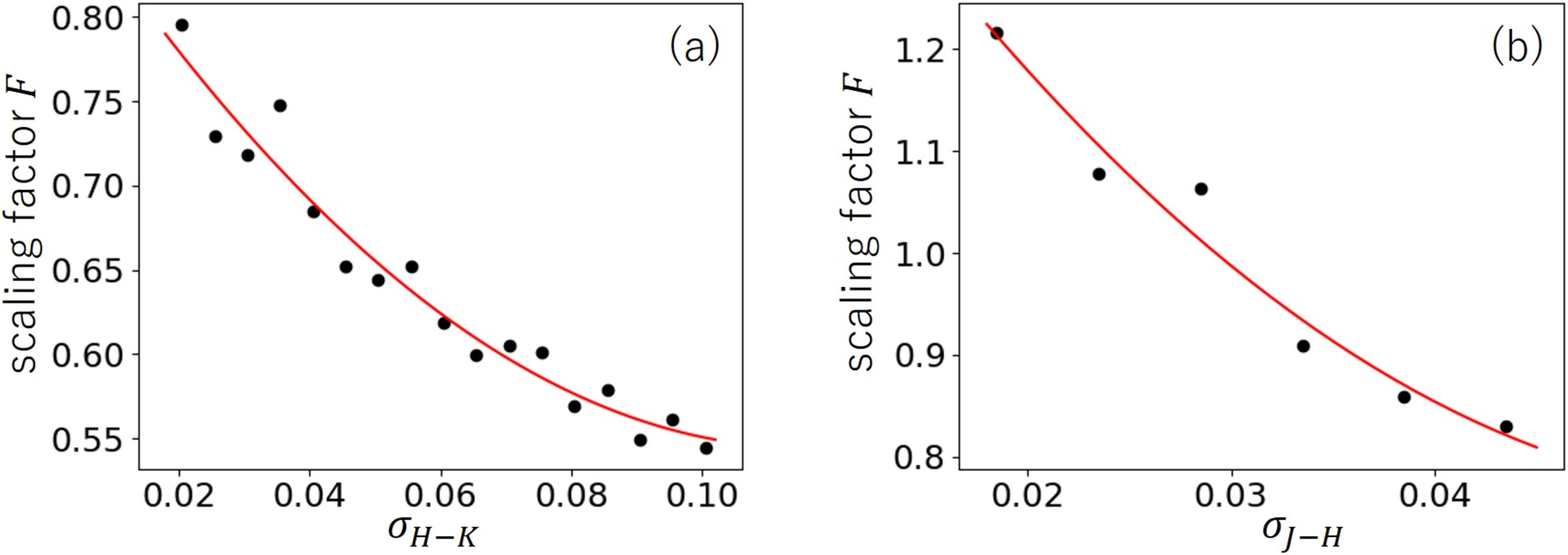} 
\end{center}
\caption{Scaling factor $F$ for the binned data of figures $\ref{fig:err_calib}$a and $\ref{fig:err_calib}$b as a function of photometric errors of (a) the $H-K$ color and (b) the $J-H$ color, respectively. Solid line represents the result of the second-order polynomial fitting.}\label{fig:scaling_factor}
\end{figure*}

More specifically, we binned the data of $\chi^2/N$ at an interval of 0.005 mag in both $\sigma_{H-K}$ and $\sigma_{J-H}$.
For each binned data at $\sigma_{H-K}=$0.018 to 0.103 mag and $\sigma_{J-H}=$0.016 to 0.046 mag, we calculate the scaling factor $F$ so that the logarithm of $\chi^2/N$ becomes zero.
Data points in figures $\ref{fig:scaling_factor}$a and $\ref{fig:scaling_factor}$b are the estimated scaling factors for each binned data of $\sigma_{H-K}$ and $\sigma_{J-H}$, respectively.
We finally perform the second-order polynomial fitting to the estimated scaling factors, and determine the scaling factor $F$ as a function of photometric errors as shown by the solid lines in figures \ref{fig:scaling_factor}a and \ref{fig:scaling_factor}b.
Figures \ref{fig:err_calib}c and \ref{fig:err_calib}d are the logarithmic plots of $\chi^2/N$ of the $H-K$ and $J-H$ colors, respectively, after applying the resultant scaling factor $F$ to the photometric errors of each star.
In the figures, we clearly see that the logarithm of $\chi^2/N$ are now closer to zero in all the range of photometric errors.
\subsubsection{Simulation of the $A_V$ scatter}\label{sec:simu_av}
In order to evaluate the $A_{V}$ scatter caused by photometric errors in each spatial bin of \timeform{5'}$\times$\timeform{5'}, we perform a Monte Carlo simulation.
We first generate a random set of colors for each star on the CC diagram in each bin assuming that the colors follow the two-dimensional Gaussian distribution, the center of which is fixed at the observed $H-K$ and $J-H$ colors.
We here adopt the corrected $\sigma_{H-K}$ and $\sigma_{J-H}$ (see sub-subsection \ref{sec:corr_error}) as the standard deviation of the Gaussian distribution.
We estimate $A_{V}$ for each simulated star and subtract observed $A_V$ from simulated $A_V$ to exclude the dust extinction.
The distribution of the simulated values are shown by the blue histogram and blue squares in figures \ref{fig:eg_percentile}b and \ref{fig:eg_percentile}c, respectively,
which can be considered as the $A_{V}$ scatter caused by photometric errors alone (i.e., without dust extinction) in the spatial bin.
We repeat this simulation 100 times and generate 100 sets of the simulated $A_V$ distribution for each spatial bin.

\subsection{Extinction along the line of sight}
\begin{figure*}
 \begin{center}
  \includegraphics[width=16cm]{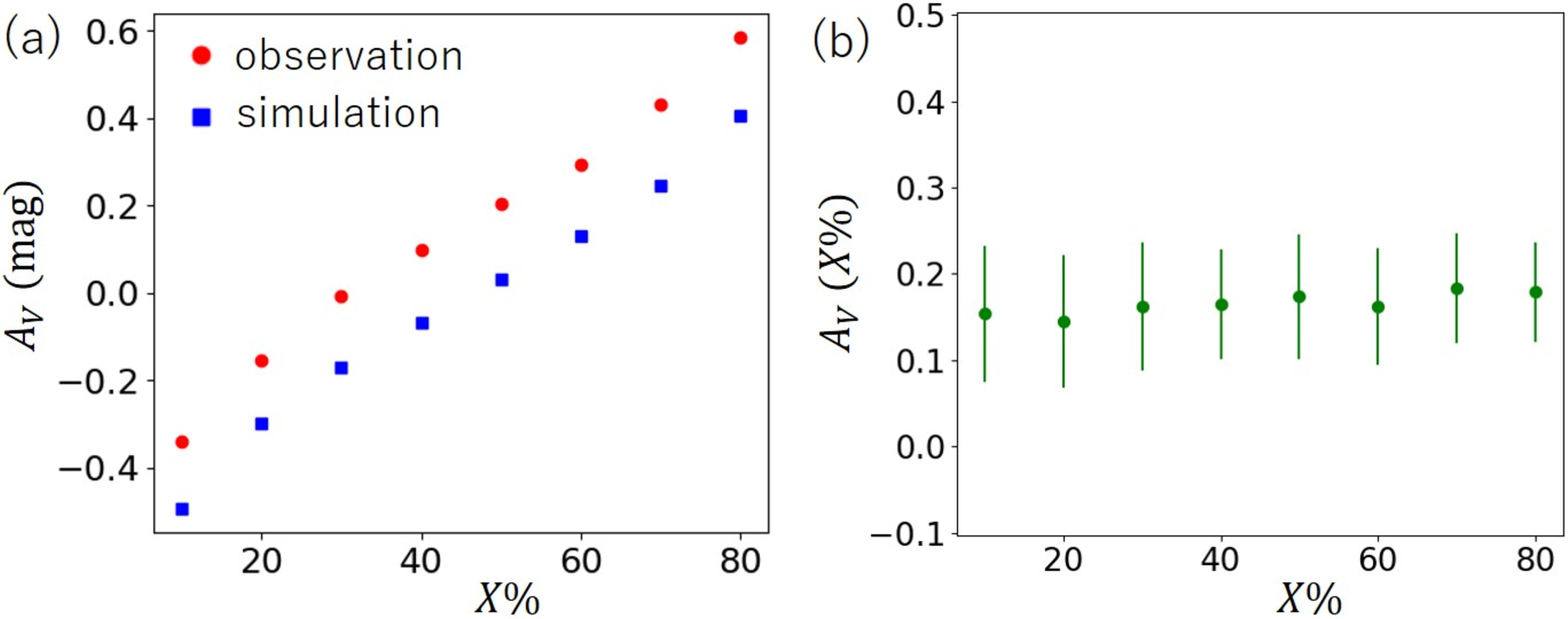} 
 \end{center}
\caption{(a) Mean $A_V$ in the range of $X$\% to $(X+10)$\% percentile of $A_V$ distribution in figure \ref{fig:eg_percentile}c for $X$=10\% to 80\% at every 10\%.
(b) Difference between observed and simulated $A_V$ at each percentile shown in panel (a).}\label{fig:eg_sim_percentile}
\end{figure*}

In order to evaluate the intrinsic $A_V$ distributions along the line of sight, we calculate the percentile values of $A_V$ based on the $X$ percentile method.
In this method, $A_V$ is originally estimated from the difference in the $X$ percentile of the color between the reference and observed field, which enables us to infer the positions of dark clouds.
In our study, we evaluate $A_V$ at each percentile by subtracting simulated $A_V$ (excluding dust extinction) from observed $A_V$ (including dust extinction).
We first average observed $A_V$ in the range of $X$\% to $(X+10)$\% percentile in figure \ref{fig:eg_percentile}c, where $X$ is 10\% to 80\% at every 10\%.
Red circles in figure \ref{fig:eg_sim_percentile}a show the resultant mean observed $A_V$ as a function of $X$\%.
We perform this procedure to the simulated $A_V$ distribution and calculate the mean percentile values of $A_V$ simulated 100 times as shown by blue squares in figure \ref{fig:eg_sim_percentile}a.
We finally calculate the difference between the observed and simulated percentile $A_V$ at each percentile as shown in figure \ref{fig:eg_sim_percentile}b.
Here and hereafter, we define $A_{V}$($X$\%) as the resultant values shown in figure \ref{fig:eg_sim_percentile}b (e.g., $A_V$(10\%) is the mean $A_V$ from 10\% to 20\% percentile of observed $A_V$ after subtraction of simulated $A_V$).
The uncertainties of $A_{V}$($X$\%) are calculated in subsection \ref{sec:err}.

The shape of the histogram of the simulated $A_V$ in figure \ref{fig:eg_percentile}b is consistent with that of the observed one, which demonstrates that our simulation using the scaling factor $F$ works well.
Indeed, $A_V$($X$\%) in figure \ref{fig:eg_sim_percentile}b is almost constant value at 0.15 mag for all $X$\%.
The constant $A_V$ is likely caused by the Galactic extinction across the LMC.
\citet{dobashi} find that the Galactic extinction is $A_V \simeq 0.2 \ \rm mag$ from the H\,\emissiontype{I} column density $N({\rm H\,\emissiontype{I}}$), assuming 
${A_V}/N({\rm H\,\emissiontype{I}})=5.34\times10^{-22}$ mag/$\rm cm^{-2}$ (\cite{galactic_dustgas}, for $R_V$=3.1).
In addition, the constant $A_{V}$ for all $X$\% in figure \ref{fig:eg_sim_percentile}b means that the dust associated with the corresponding region of the LMC is not detected significantly along the line of sight as expected from the low column density region.

%
%
\subsection{Uncertainties of dust extinction}\label{sec:err}
We first calculate the $A_V$ uncertainty of each star.
When we perform the Monte Carlo simulation as described in subsection \ref{sec:sim_av}, the $A_V$ value of each star is estimated 100 times.
We estimate the $A_V$ uncertainty of each star as the standard deviation of $A_V$ simulated 100 times.
As a next step, we calculate the uncertainties in the mean values of $A_V$ included in  $X$\% to $(X+10)$\% percentile, $\sigma_{A_{V,X}}$, as
\begin{equation}
\sigma_{A_{V,X}}=\frac{1}{N_X} \sqrt{\sum_{i=1}^{N_X} \delta A_{Vi}^2},
\label{eq:2d_err}
\end{equation}
where $N_X$ and $\delta A_{Vi}$ are the number of stars included in $X$\% to $(X+10)$\% percentile, and the $A_V$ uncertainty of the $i$th star, respectively.
$\sigma_{A_{V,X}}$ for each $X$\% is shown in figure \ref{fig:eg_sim_percentile}b as error bars.

When we compare $A_V$($X$\%) with the hydrogen column density to evaluate detailed dust geometry (section \ref{sec:comp_gas}), we additionally consider the uncertainty of the percentile virtually corresponding to the uncertainty of the distance along the line of sight by using a simple count uncertainty (i.e., $\sqrt{N}$).
We calculate its uncertainty as $\sigma_{{d},X}=(A_{Vk} - A_{Vl})/2$, where $A_{Vk}$ and $A_{Vl}$ are the $A_V$ values for the stars of the index of $k$ and $l$ defined as the integer part of ($i_{\rm{center}}+\sqrt{N}$) and ($i_{\rm{center}}-\sqrt{N}$), respectively.
Here, $i_{\rm center}$ is the median index of stars included in $X$\% to $(X+10)$\% percentile, and $N$ is the total number of stars in the spatial bin.
We finally calculate the total uncertainties of $A_V$($X$)\% from the error propagation, $\delta A_V(X\%)=\sqrt{\sigma_{A_{V,X}}^2 + \sigma_{{d},X}^2}$.

\section{Result}
\subsection{Results for the individual lines of sight}
\begin{figure*}
 \begin{center}
  \includegraphics[width=16cm]{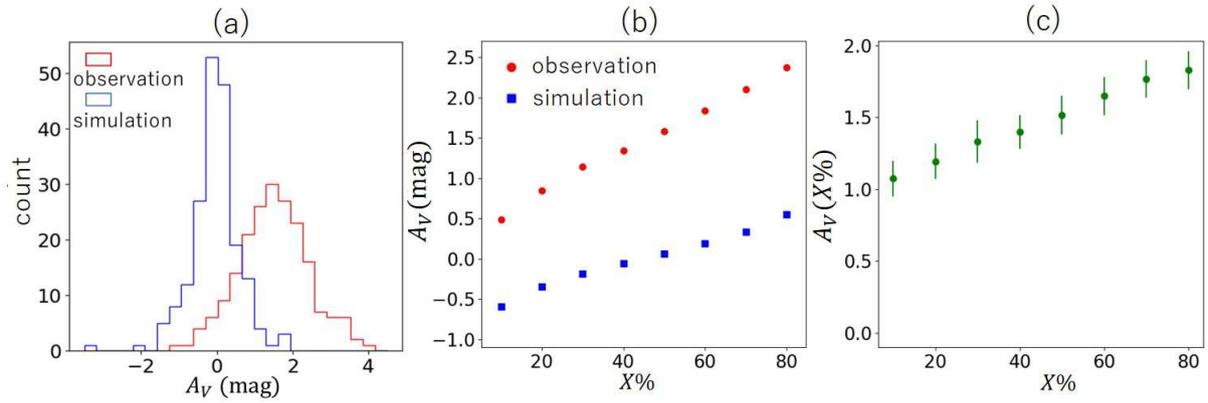} 
 \end{center}
\caption{(a) Histogram of observed (red) and simulated (blue) $A_V$ for the spatial bin of \timeform{5'}$\times$\timeform{5'} centered at $(\alpha, \delta)_{\rm J2000.0} =$ (\timeform{5h38m}, $-$\timeform{69D10'}) in 30 Dor. Panels (b) and (c) are the same diagram as in figures \ref{fig:eg_sim_percentile}a and \ref{fig:eg_sim_percentile}b, respectively, but corresponding to panel (a).}\label{fig:res_eg1}
\end{figure*}

\begin{figure*}
 \begin{center}
  \includegraphics[width=16cm]{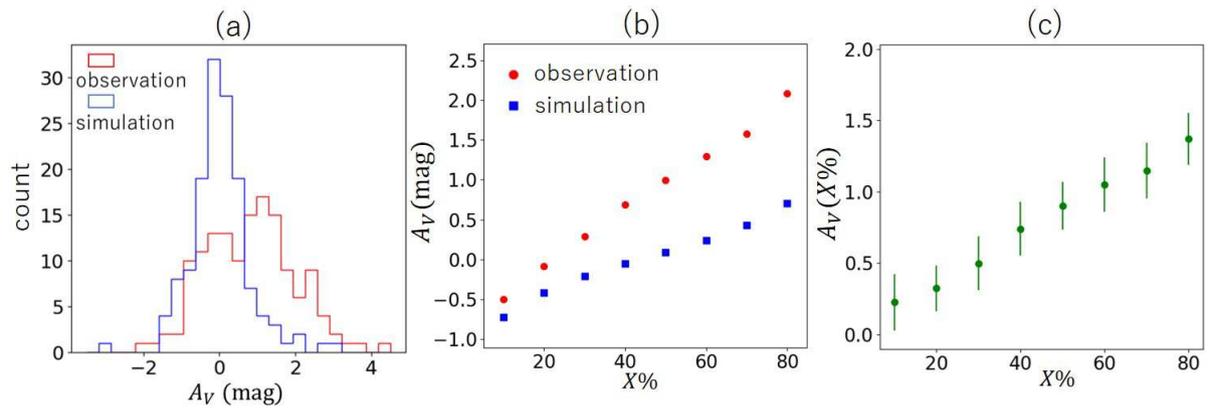} 
 \end{center}
\caption{Same diagrams as in figure \ref{fig:res_eg1} but for the spatial bin of \timeform{5'}$\times$\timeform{5'} centered at $(\alpha, \delta)_{\rm J2000.0} =$ (\timeform{5h32m}, $-$\timeform{68D28'}).}\label{fig:res_eg2}
\end{figure*}

The results for two example lines of sight are presented in figures \ref{fig:res_eg1} and \ref{fig:res_eg2} for the spatial bins of \timeform{5'}$\times$\timeform{5'} centered at $(\alpha, \delta)_{\rm J2000.0} =$ (\timeform{5h38m}, $-$\timeform{69D10'}) in 30 Dor and (\timeform{5h32m}, $-$\timeform{68D28'}), respectively (the positions of these regions are plotted in figures \ref{fig:av_w_gas} and \ref{fig:tau_w_av}).
In the former region, gas collision is suggested and thus the dust cloud is expected to be located in front of the LMC disk (\cite{furuta}).
In the latter region, there is relatively high column density gas of the LMC disk (see figure \ref{fig:av_w_gas}c), and the gas collision is not suggested.

In figure \ref{fig:res_eg1}a, we recognize that the observed $A_{V}$ histogram is shifted to positive values and broader compared to the simulated one.
$A_V$(10\%) of about 1.1 mag in figure \ref{fig:res_eg1}c suggests that the dust cloud is located in front of almost all stars as expected from the gas colliding region and largely contribute to the dust extinction in this region.
In addition, the broader $A_{V}$ histogram is likely to be caused by the diffuse extinction existing along the line of sight.
The former trend is also suggested by \citet{tatton} who estimated $A_V$ for each star around 30 Dor using a CM diagram ($J-K$ vs. $K$) and found that almost all the stars around the region have large dust extinction ($A_V > 1.2$ mag).

In figure \ref{fig:res_eg2}a, we find two peaks around $A_V$=0.0 mag and 1.0 mag in the observational histogram.
$A_V$(10\%) of 0.2 mag in figure \ref{fig:res_eg2}c is consistent with the Galactic foreground extinction, which suggests that the first peak in the $A_V$ histogram is attributed to stars not affected by the dust extinction in the LMC itself.
On the other hand, the second peak of the $A_V$ histogram is likely attributed to stars behind the dust cloud in the LMC, causing a rapid increase of $A_V$ at $X$=30\% and onward in figure \ref{fig:res_eg2}c.
This result is consistent with the existence of the high column density gas of the LMC disk.

%
%
\subsection{Extinction map along the line of sight}
\begin{figure*}
 \begin{center}
  \includegraphics[width=16cm]{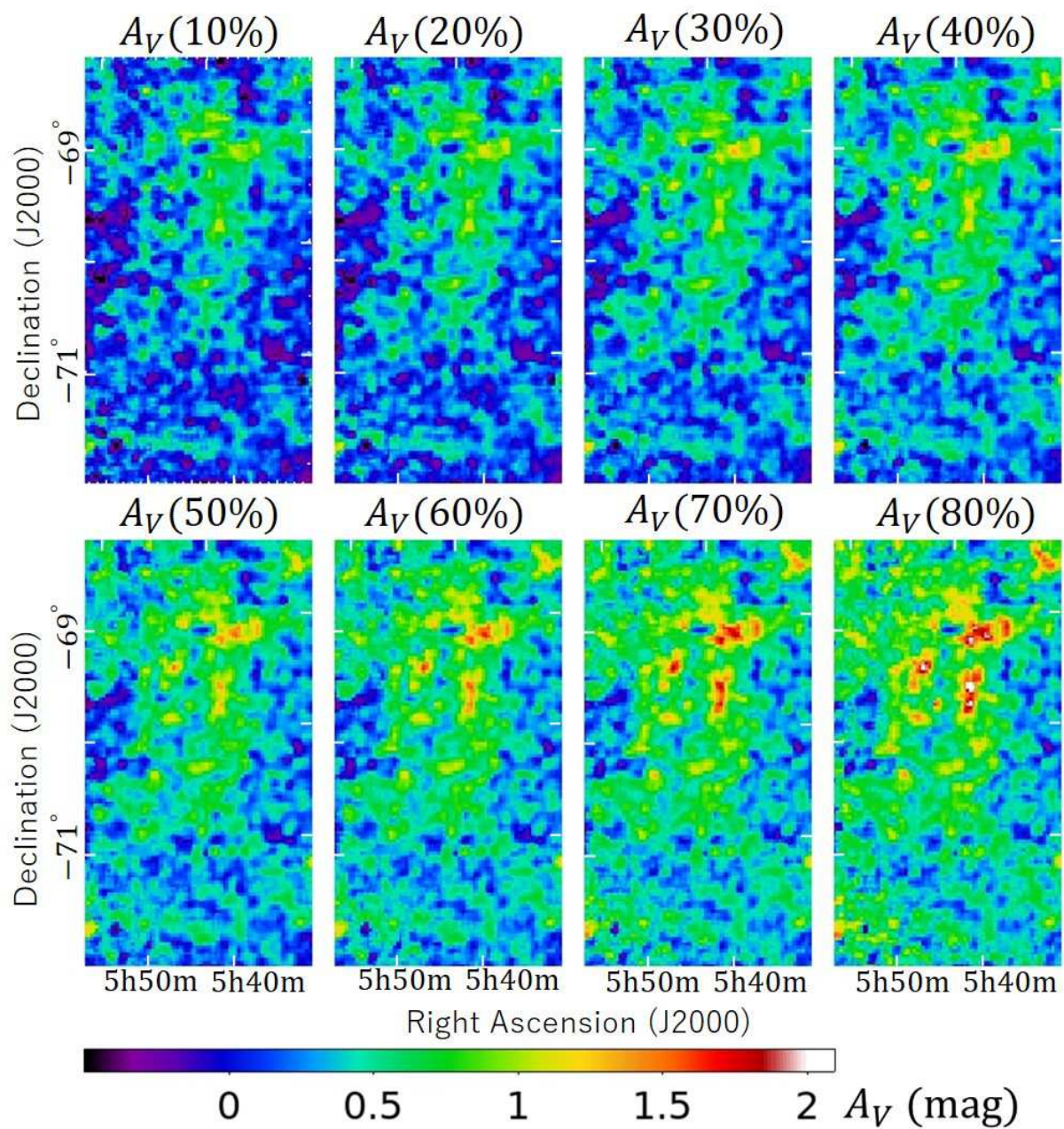} 
 \end{center}
\caption{Cumulative dust extinction map from an observer to stars included in $X$\% to $(X+10)$\% percentile for $X$=10\% to 80\% at every 10\%. The angular resolution of the maps is \timeform{5'}.}\label{fig:cumulative_map}
\end{figure*}

\begin{figure*}
 \begin{center}
  \includegraphics[width=16cm]{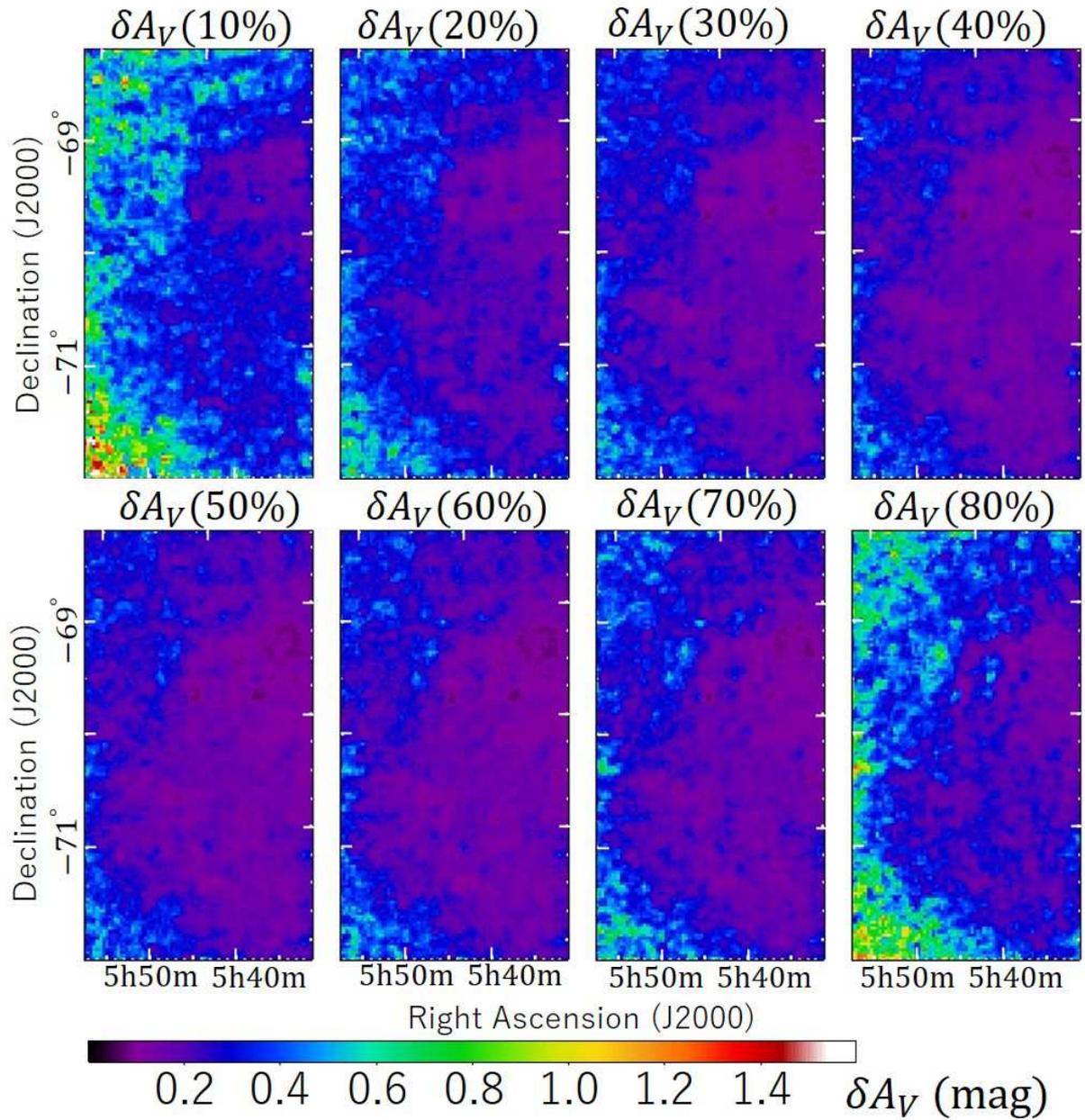} 
 \end{center}
\caption{Uncertainties of the dust extinction maps shown in figure \ref{fig:cumulative_map}.}\label{fig:errmap}
\end{figure*}

In this subsection, we consider large-scale dust geometry.
Figure \ref{fig:cumulative_map} shows the $A_V$(10\%) to $A_V$(80\%) maps in the H\,\emissiontype{I} ridge region.
As described in section \ref{sec:method}, $A_V$ is calculated with \timeform{5'}$\times$\timeform{5'} bins every \timeform{1'.6} grid and thus the PSF of the $A_V$ map is of a rectangular shape with the spatial resolution of \timeform{5'} and the grid size of \timeform{1'.6}.
Since $A_V$($X$\%) is the cumulative dust extinction from an observer to $X$\%, $A_V$ increases with $X$\% as can be seen in the figure.
The $A_V$(80\%) map is considered to trace almost all the dust in the LMC.
We present the uncertainty map of each $A_V$ map in figure \ref{fig:errmap}.
The uncertainty is relatively large in the east of the H\,\emissiontype{I} ridge region due to the lower density of stars (see figure \ref{fig:number_density}).

In order to evaluate the Galactic foreground extinction using the derived dust extinction map, we calculate the averaged $A_V$ in the $A_V$(10\%) map where the total hydrogen column density $N({\rm H})$ is lower than $5.0 \times 10^{20} \ \rm cm^{-2}$ (the total hydrogen column density is defined in the next subsection).
As a result, we obtain 0.15 mag, which is consistent with the foreground extinction towards the LMC ($\sim$ 0.2 mag) estimated by \citet{dobashi}.

\begin{figure*}
 \begin{center}
  \includegraphics[width=16cm]{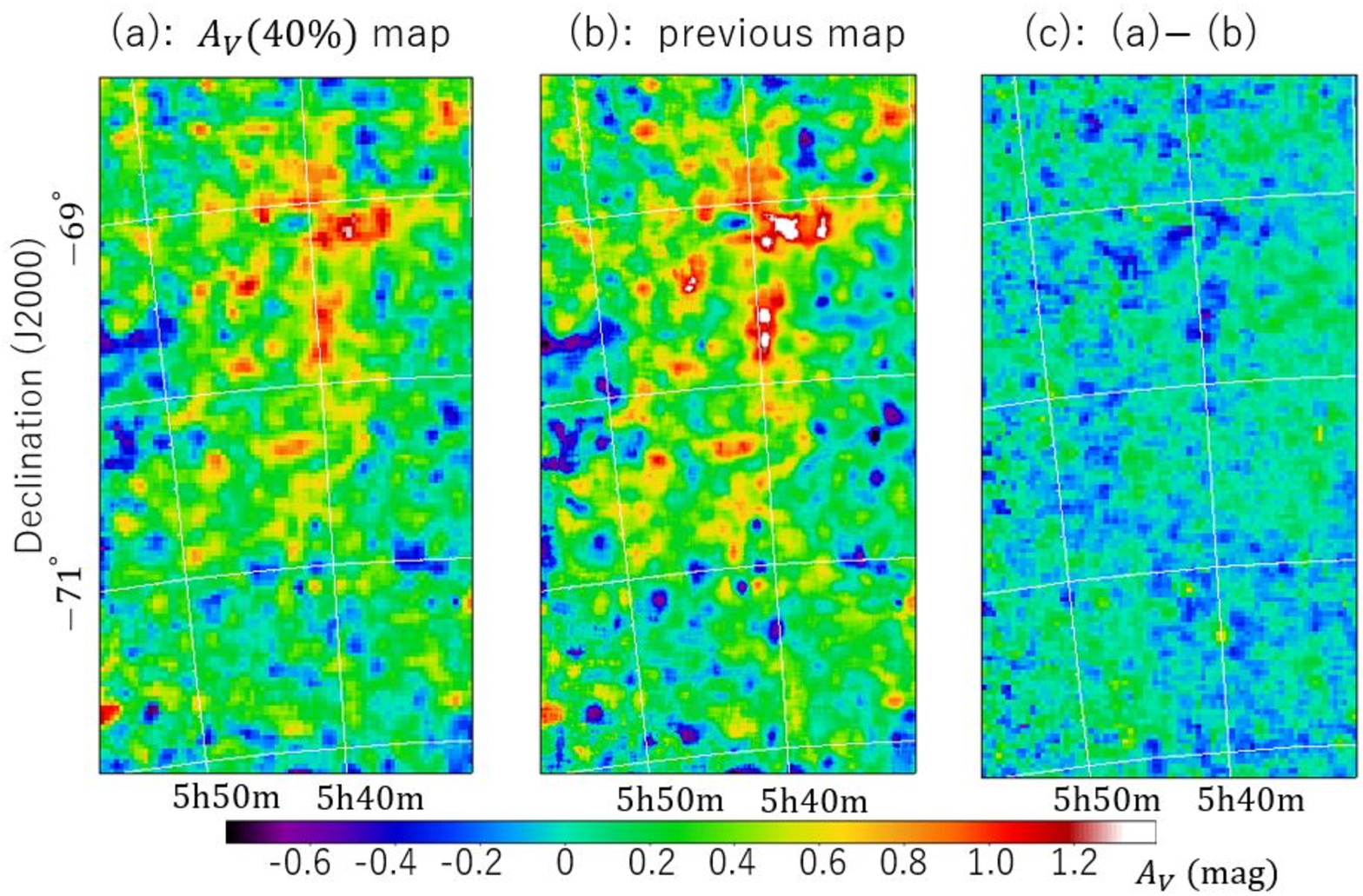} 
 \end{center}
\caption{(a) $A_V$(40\%) map derived in this paper. (b) Mean $A_V$ map derived from \citet{furuta}. The Galactic foreground extinction is subtracted in both panels (a) and (b). (c) Difference in $A_V$ between panels (a) and (b).}\label{fig:comp_pre}
\end{figure*}

We compare the derived $A_V$ map with the previous one in \citet{furuta}.
Since the previous $A_V$ map was estimated from the mean observed colors, we use the $A_V$(40\%) map (i.e., mean $A_V$ from 40\% to 50\% percentile $A_V$) for comparison  after subtracting the Galactic foreground extinction. 
Figures \ref{fig:comp_pre}a and \ref{fig:comp_pre}b show the $A_V$(40\%) map and our previous $A_V$ map, respectively, whose spatial resolution is reduced to be the same as that of the $A_V$(40\%) map.
Figure \ref{fig:comp_pre}c shows the $A_V$(40\%) map after subtraction of the previous map, from which we confirm that the $A_V$(40\%) map is consistent with the previous $A_V$ map, although local differences can be seen at high extinction regions.
\section{Comparison of dust extinction with $N$(H)}\label{sec:comp_gas}
\subsection{Gas tracer}
As an atomic gas tracer, we use the H\,\emissiontype{I} velocity-integrated intensity maps constructed by \citet{fukui_2017} and \citet{tsuge}, where the Australia Telescope Compact Array (ATCA) and Parkes H\,\emissiontype{I} $21\ \rm cm$ data (\cite{himap}) are used.
They subtracted the galactic rotation velocity from the H\,\emissiontype{I} data, which is defined as $V_{\rm offset}$, and decomposed the H\,\emissiontype{I} map into three velocity components (see \cite{tsuge} for details of the analysis).
These velocity components are defined as the L-, I- and D-components with the integrated velocity ranges of $V_{\rm offset}=$ $-$100 to $-$30, $-$30 to $-$10 and $-$10 to 10 $\rm km\ s^{-1}$, respectively.
A typical H\,\emissiontype{I} spectrum in the H\,\emissiontype{I} ridge region is shown in figure 1 of \citet{tsuge} together with the velocity range of each velocity component (see also figure 8 of \cite{tsuge_2021} for examples of the H\,\emissiontype{I} spectra in the LMC H\,\emissiontype{I} ridge region).
The L-, I- and D-components are named after the initial letters of ``Low'', ``Intermediate'' and ``Disk'' velocity, respectively.
As a molecular gas tracer, we use the rotational transitions of $^{12}{\rm CO}\ (J=$1--0) observed with the NANTEN 4 $\rm m$ telescope (\cite{comap}).
The CO integrated intensity map is decomposed into the L-, I- and D-components according to the above velocity ranges of the H\,\emissiontype{I} integrated intensity maps.

The H\,\emissiontype{I} integrated intensity  map is converted into the H\,\emissiontype{I} column density map by using the conversion factor, $X_{\rm H\,\emissiontype{I}}=1.82 \times 10^{18}\ \rm{cm^{-2}}/({\rm K\ km\ s^{-1}})$ (\cite{nh_conv}). 
CO is used to trace $\rm H_{2}$ through the CO-to-$\rm H_{2}$ conversion factor, $X_{\rm CO}$.
We reduce the spatial resolutions of the H\,\emissiontype{I} column density maps and CO maps to be the same as that of the $A_{V}$ map at a resolution of \timeform{5'} by convolving with a boxcar kernel of \timeform{5'}$\times$\timeform{5'}.
Thus the PSF of the gas maps is of the same rectangular shape with that of the $A_V$ map to make a comparison on a pixel-by-pixel basis (subsection \ref{sec:comp_av_gas}).
We calculate the total gas column density $N({\rm H})$ of each of the L-, I- and D-component as
\begin{equation}
N({\rm H})\ =\ N({\rm H\,\emissiontype{I}})\ + \ 2N({\rm H_{2}}),
\label{nh}
\end{equation}
where $N({\rm H\,\emissiontype{I}})$ and $N({\rm H_2})$ are the H\,\emissiontype{I} and the $\rm H_{2}$ column densities, respectively.

\subsection{$A_{V}$ vs. $N$(H) correlation}
\begin{figure*}
 \begin{center}
  \includegraphics[width=16cm]{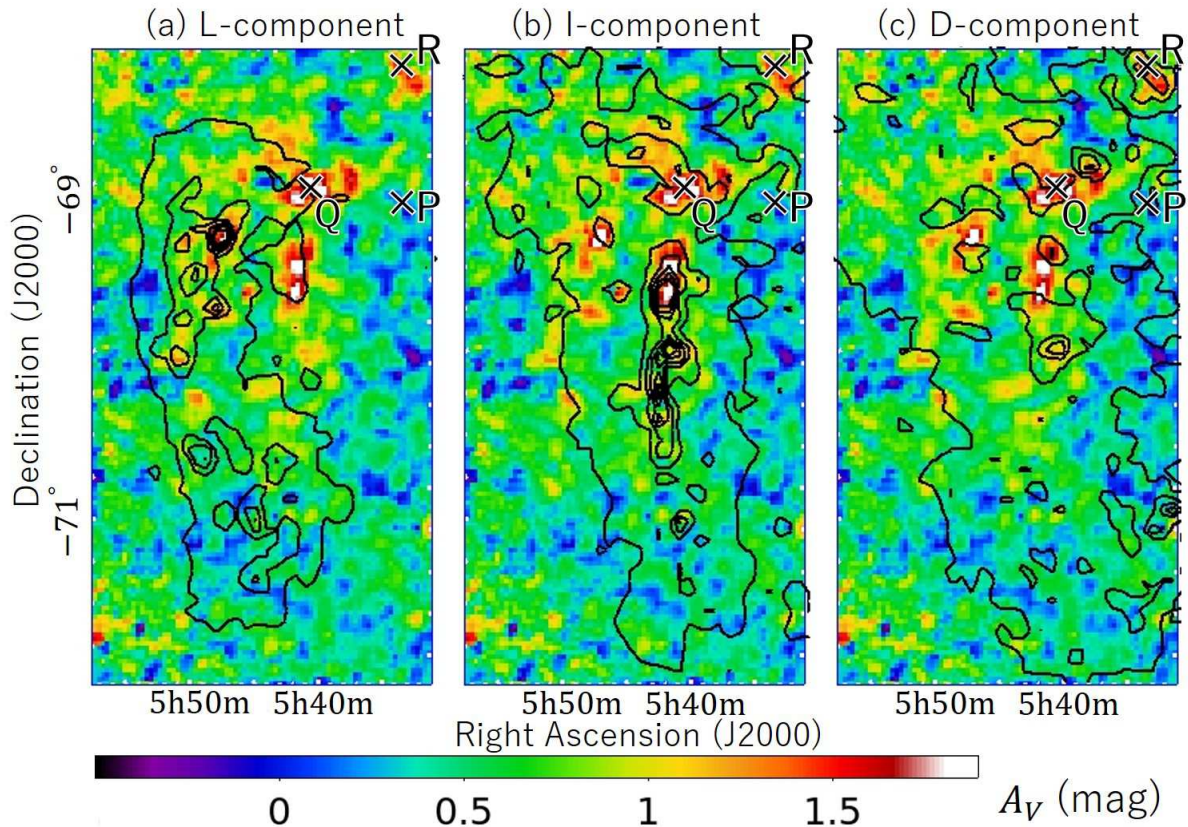} 
 \end{center}
\caption{$A_V$(80\%) map superposed on the $N(\rm H)$ contour maps of the (a) L-, (b) I- and (c) D-component assuming $X_{\rm CO}=7\times10^{20}\ {\rm cm^{-2}}/(\rm{K\ km\ s^{-1}})$. The contour levels are (0.6, 2.1, 3.6, 5.0, 6.5 and 8.0)$\times 10^{21}\ {\rm cm^{-2}}.$ The cross symbols labeled ``P'', ``Q'' and ``R'' show the positions of the spatial bins where the plots in figures \ref{fig:eg_percentile}, \ref{fig:res_eg1} and \ref{fig:res_eg2}, respectively, are created.
}\label{fig:av_w_gas}
\end{figure*}

To check the spatial correlation between $A_V$ and $N(\rm H)$, we present the $A_V$(80\%) maps with the contours of $N(\rm H)$ of the L-, I- and D-components in figure \ref{fig:av_w_gas}.
Here, we adopt $X_{\rm CO}=7\times10^{20}\ {\rm cm^{-2}}/(\rm{K\ km\ s^{-1}})$ which is the averaged $X_{\rm CO}$ in the LMC derived from \citet{h2_conv}. 
Because $N(\rm H)$ is the integrated value along the line of sight, the $A_V$(80\%) map is expected to correlate with the $N(\rm H)$ maps best among the $A_V$(10\%) to $A_V$(80\%) maps. 
Indeed, the I- and D-component gases correlate well with the dust extinction.
On the other hand, the L-component correlates well with the dust extinction in the northern part but not in the southern part, as can be seen from figure \ref{fig:av_w_gas}a.
This trend can also be seen in \citet{furuta}, who consider the gas distribution based on the scenario that the L-component gas is falling into the LMC disk and the collision is on-going in the H\,\emissiontype{I} ridge region (\cite{fukui_2017}).

\subsection{Decomposition of $A_V$ into different velocities}\label{sec:comp_av_gas}
We estimate the fractions of $A_V$ due to the dust associated with the L-, I- and D-components for every percentile from 10\% to 80\%, applying the fitting method described in \citet{furuta} to our newly constructed $A_V$ maps.
The fitting function is expressed as the following equation:
\begin{eqnarray}\label{eq:regress1}
A_V = \theta(y- y_0)\ a N({\rm H})_{\rm L}\ + \ b N({\rm H})_{\rm I}\ + \ c N({\rm H})_{\rm D}, \\ 
\theta(y- y_0) = \left\{ \begin{array}{ll}
    1 \ & (y\ge y_0), \\
    0 \ & (y < y_0), \\
  \end{array} \right. \nonumber \\
N({\rm H})_k = N({\rm H}\,\emissiontype{I})_{k} + 2x_{k} W({\rm CO})_{k} \ &  (k=\rm L,\ I\ or\ D), \nonumber
\end{eqnarray}
where $a$, $b$ and $c$ are the free parameters indicating the dust/gas ratios (i.e., $A_V$/$N$(H)) of the L-, I- and D-components, respectively, while $N({\rm H})_k$ for $k$=L, I or D is the gas column density of the L-, I- or D-component.
Here, $y$ and $y_0$ are the pixel coordinates along the Declination direction.
$\theta (y-y_0)$ is a step function and the free parameter $y_0$ indicates that the L-component at $y >y_0$ is located in front of the LMC disk, contributing to the dust extinction, whereas the L-component at $y < y_0$ is located behind the LMC disk, not contributing to the dust extinction (see figure 7 of \cite{furuta}).
For simplicity, we assume that the boundary is parallel to the Right Ascension direction.
$N({\rm H}\,\emissiontype{I})_{k}$ and $ W({\rm CO})_{k}$ are the ${\rm H}\,\emissiontype{I}$ column density and the integrated CO intensity of each velocity component, respectively.
$x_{k}$ is the free parameter corresponding to $X_{\rm CO}$ of the L-, I- or D-component. 
We set the $X_{\rm CO}$ factor of each velocity component as a free parameter because \citet{furuta} revealed that $X_{\rm CO}$ of each velocity component is significantly different from each other.

As demonstrated by \citet{furuta}, we can decompose $A_V$ into the three different velocity components with the linear regression of equation (\ref{eq:regress1}) using $\delta A_V(X\%)$ calculated in subsection \ref{sec:err}.
In this study, we newly apply this method to each $A_V$($X$\%) map after subtraction of the Galactic foreground extinction, which enables us to decompose $A_V$ existing along the line of sight into different velocity components.
In other words, we can evaluate the three-dimensional dust geometry associated with the L-, I- and D-components.
We discuss the dust geometry and dust/gas ratios of them in the next section.

When we perform the linear regression, we use only the region where $N({\rm H})_{\rm D}$ is higher than $1.0 \times 10^{20} \ \rm cm^{-2}$ assuming  $X_{\rm CO}=7\times10^{20}\ {\rm cm^{-2}}/(\rm{K\ km\ s^{-1}})$.
In addition, we mask the 30 Dor region where the surface brightness is higher than $20\ \rm MJy/sr$ in the Spitzer/MIPS $24\ \rm \mu m$ map from the SAGE program (\cite{mips}), because \citet{30dor} suggested that the reddening law around the 30 Dor region does not follow that of \citet{red_law2}, which causes the under-estimation of the dust/gas ratio around 30 Dor.

\begin{figure*}
 \begin{center}
  \includegraphics[width=16cm]{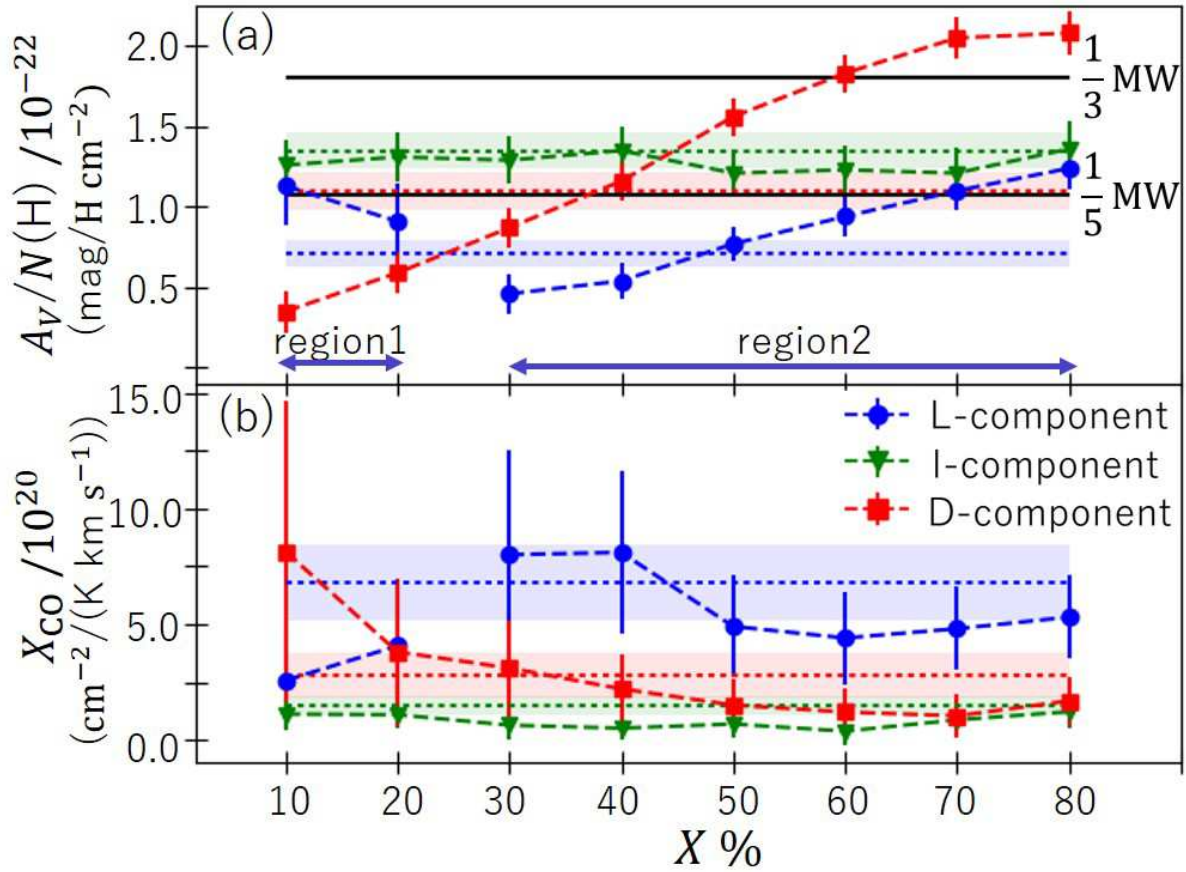} 
 \end{center}
\caption{Fitting parameters of (a) $A_V$/$N$(H) and (b) $X_{\rm CO}$ derived from the linear regression of equation (\ref{eq:regress1}) as a function of $X$\%.
Circles, triangles and squares are the parameters derived for the L-, I- and D-components, respectively.
Data points and dotted lines are color-coded according to their velocity components.
Dotted horizontal lines and shaded regions show the parameter of each velocity component and 1$\sigma$ level spread around the parameter, respectively, derived from \citet{furuta}.
In panel (a), solid horizontal lines correspond to 1/3 and 1/5 of $A_V$/$N$(H) in the Milky Way (MW). ``region 1'' and ``region 2'' are part of the L-component separated by the derived boundary positions of $y_0$ (see subsection \ref{sec:discuss}).}\label{fig:fit_res}
\end{figure*}

The resultant fitting parameters ($y_0$, $a$, $b$, $c$ and $x_k$) are summarized in table \ref{tab:fit_res}.
The uncertainties of the free parameters are estimated by the formal regression errors using the $A_V$ uncertainties.
We plot $A_V$/$N$(H) and $X_{\rm CO}$ thus obtained for each velocity component and those derived from \citet{furuta} as a function of $X$\% in figures \ref{fig:fit_res}a and \ref{fig:fit_res}b, respectively. 
\begin{table*}
\tbl{Fitting parameters derived from comparison of the dust extinction and $N(\rm H)$ of each component.}{
 \begin{tabular}{lcccccccc}
\hline
\multicolumn{1}{c}{Name} &Reduced $\chi^2$ &$y_0$\footnotemark[*]& $a$ $\left( \frac{A_V}{N(\rm H)} \right)$\footnotemark[$\dagger$] &$x_{\rm L}$ $\left(X_{\rm CO}\right)$\footnotemark[$\ddagger$] &$b$ $\left( \frac{A_V}{N(\rm H)} \right)$\footnotemark[$\dagger$] &$x_{\rm I}$ $\left(X_{\rm CO}\right)$\footnotemark[$\ddagger$] & $c$ $\left( \frac{A_V}{N(\rm H)} \right)$\footnotemark[$\dagger$] &$x_{\rm D}$ $\left(X_{\rm CO}\right)$\footnotemark[$\ddagger$] \\
 & & & (L-component) & (L-component) & (I-component) & (I-component) & (D-component) & (D-component)\\
 \hline
$A_V$(10\%) & 0.51 & \timeform{-69D.4}$\pm$\timeform{0D.1} & $1.13\pm 0.24$ &$2.6\pm 2.1$ & $1.26\pm 0.16$ &$1.1\pm 0.7$ &  $0.35\pm 0.13$ &$8.1\pm 6.5$ \\
$A_V$(20\%) & 0.91 & \timeform{-69D.4}$\pm$\timeform{0D.1} & $0.91\pm 0.23$ &$4.1\pm 2.6$ & $1.31\pm 0.15$ &$1.1\pm 0.6$ &  $0.59\pm 0.13$ &$3.8\pm 3.2$ \\
$A_V$(30\%) & 1.21 & \timeform{-70D.9}$\pm$\timeform{0D.2} & $0.46\pm 0.12$ &$8.0\pm 4.5$ & $1.29\pm 0.15$ &$0.6\pm 0.6$ &  $0.87\pm 0.12$ &$3.1\pm 2.1$ \\
$A_V$(40\%) & 1.37 & \timeform{-70D.9}$\pm$\timeform{0D.2} & $0.54\pm 0.11$ &$8.1\pm 3.5$ & $1.35\pm 0.15$ &$0.5\pm 0.5$ &  $1.16\pm 0.12$ &$2.2\pm 1.5$ \\
$A_V$(50\%) & 1.44 & \timeform{-70D.9}$\pm$\timeform{0D.2} & $0.77\pm 0.11$ &$4.9\pm 2.2$ & $1.21\pm 0.15$ &$0.7\pm 0.6$ &  $1.56\pm 0.12$ &$1.5\pm 1.1$ \\
$A_V$(60\%) & 1.40 & \timeform{-70D.9}$\pm$\timeform{0D.2} & $0.94\pm 0.12$ &$4.4\pm 2.0$ & $1.23\pm 0.15$ &$0.4\pm 0.6$ &  $1.83\pm 0.12$ &$1.2\pm 1.0$ \\
$A_V$(70\%) & 1.26 & \timeform{-70D.9}$\pm$\timeform{0D.2} & $1.10\pm 0.12$ &$4.8\pm 1.8$ & $1.21\pm 0.16$ &$0.9\pm 0.7$ &  $2.05\pm 0.13$ &$1.0\pm 0.9$ \\
$A_V$(80\%) & 0.80 & \timeform{-70D.9}$\pm$\timeform{0D.2} & $1.24\pm 0.13$ &$5.3\pm 1.8$ & $1.36\pm 0.17$ &$1.2\pm 0.7$ &  $2.08\pm 0.14$ &$1.7\pm 1.0$ \\
previous map\footnotemark[$\$$] & 1.16 & \timeform{-70D.8}$\pm$\timeform{0D.2} & $0.71\pm 0.08$ &$6.8\pm 1.6$ & $1.35\pm 0.11$ &$1.5\pm 0.4$ &  $1.10\pm 0.12$ &$2.8\pm 1.0$ \\
 \hline
 \end{tabular}}\label{tab:fit_res}
 \begin{tabnote}
 \footnotemark[*] The boundary position of $y_0$ in the units of pixel is converted into $\delta_{\rm J2000.0}$ at $\alpha_{\rm J2000.0} = $\timeform{87D.4}\\
  \footnotemark[$\dagger$] Given in the units of $10^{-22}$ mag/(H $\rm cm^{-2}$)\\
  \footnotemark[$\ddagger$] Given in the units of $10^{20}\ {\rm cm^{-2}}/(\rm{K\ km\ s^{-1}}$)\\
 \footnotemark[$\$$] \citet{furuta} allowing the $X_{\rm CO}$ factors of the L-, I- and D-components to vary.
\end{tabnote}
 \end{table*}
\section{Discussion}
\subsection{Large-scale dust geometry}\label{sec:discuss}
\begin{figure*}
 \begin{center}
  \includegraphics[width=16cm]{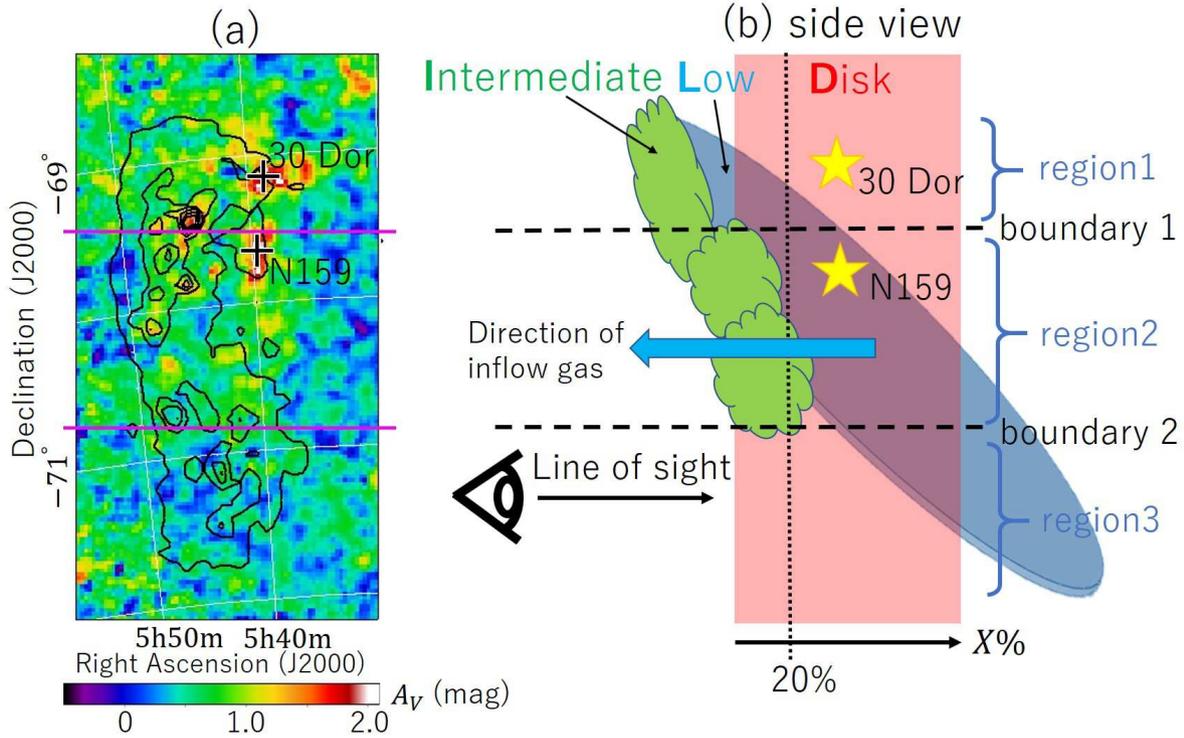} 
 \end{center}
\caption{(a) $A_V$(80\%) map superposed on the same contours of the L-component as in figure \ref{fig:av_w_gas}a. Magenta horizontal lines show the estimated boundary positions of the L-component. (b) An illustration of the side view of the expected distributions of the L-, I- and D-components.} Red rectangle and blue disk denote the LMC gas (D-component) and an inflow gas from outside the LMC (L-component), respectively. Green clouds are the I-component affected by the interaction between the L- and D-components.\label{fig:boundary_res}
\end{figure*}

We consider the dust geometry of the L, I and D-components, based on the results of the linear regression fitting.
As shown in table \ref{tab:fit_res}, the estimated boundary positions of $y_0$ at $\alpha_{\rm J2000.0} = $\timeform{87D.4} are $\delta_{\rm J2000.0} =$ \timeform{-69D.4}$\pm$\timeform{0D.1} for the maps of $A_V$(10\%) and $A_V$(20\%), while $\delta_{\rm J2000.0} =$ \timeform{-70D.9}$\pm$\timeform{0D.2} for the maps of $A_V$(30\%) to $A_V$(80\%), both of which are denoted by the horizontal lines in figure \ref{fig:boundary_res}a.
Here, we name the former and latter boundaries boundary 1 and boundary 2, respectively, and the three regions separated by boundaries 1 and 2 as regions 1 to 3 (figures \ref{fig:fit_res}a and \ref{fig:boundary_res}b).
The position of boundary 2 is consistent with that obtained by \citet{furuta} using the averaged $A_V$ map (see table \ref{tab:fit_res}).
We newly find the presence of boundary 1 owing to our new method evaluating the dust distribution along the line of sight.
The resultant two boundary positions indicate that the L-component in region 1 is located from observers to $X$=20\%, while the L-component in region 2 is located on the far side from $X$=20\%.
In addition, the L-component in region 3 is located behind the LMC.

\citet{furuta} only revealed that the L-component in region 3 is behind the LMC disk.
The present study enables us to evaluate the three-dimensional geometry of the L-, I- and D-components using $A_V$/$N(\rm H)$ for each $X$\%.
Note that $A_V$ in figure \ref{fig:fit_res}a is an integrated value towards $X\%$ (i.e., variable depending on $X$\%), while $N(\rm H)$ is a constant value integrated over the entire range along the line of sight.
Thus, in principle, $A_V$/$N(\rm H)$ increases more or less with $X$\% for a given line of sight.
However, since the decomposition of $A_V$ to each velocity component is determined by the fitting, $A_V$/$N(\rm H)$ can decrease with $X$\% (e.g., the decrease in $A_V/N$(H) of the L-component within the errors from $X$=10\% to 20\%).

Using $A_V$/$N(\rm H)$ for each $X$\% and the boundary positions of the L-component, we suggest the geometry of each velocity component as illustrated schematically in figure \ref{fig:boundary_res}b.
First, $A_V$/$N(\rm H)$ of the D-component monotonically increases with $X$\% in figure \ref{fig:fit_res}a, which indicates that the D-component (i.e., Disk component) is extended along the line of sight as shown by the red rectangle in figure \ref{fig:boundary_res}b.
This trend is consistent with a general hypothesis that stars in the LMC are mixed with the LMC gas.
On the other hand, $A_V$/$N(\rm H)$ of the I-component is constant within the errors over the integration range in figure \ref{fig:fit_res}a, which implies that dust in the I-component is located in front of almost all the stars in the LMC as shown by the green clouds in figure \ref{fig:boundary_res}b.

$A_V$/$N(\rm H)$ of the L-component from $X$=30\% to 80\% increases with $X$\% as can be seen in figure \ref{fig:fit_res}a, which indicates that the L-component in region 2 is extended from $X=30$\% to $X=80$\% .
On the other hand, $A_V$/$N(\rm H)$ of the L-component at $X$=10\% and 20\% is constant within the errors, which is the same trend as the I-component, indicating that region 1 is located in front of almost all the stars.
The significant decrease in $A_V$/$N(\rm H)$ of the L-component from $X=20\%$ to 30\% is caused by the change of $y_0$ in equation \ref{eq:regress1}, i.e., the difference of the $N$(H) distribution used for the fitting.
The expected geometry of the L-component is shown by the blue disk in figure \ref{fig:boundary_res}b, suggesting that the L-component is penetrating the LMC disk at region 2, and is decelerating through interaction with the D-component, which causes the I-component (i.e., intermediate velocity) located in front of almost all the stars in the LMC.
30 Dor and N159 are expected to be located behind the I-component as shown in figure \ref{fig:boundary_res}b, whereas 30 Dor and N159 in figure \ref{fig:av_w_gas}b show good spatial correlation with the I-component.
Since the I-component is considered to be evidence for gas collision, their spatial correlation supports that the massive star formation in these region may have been triggered by collisions of the LMC disk with clouds as seen in the I-component (\cite{fukui_2017}, \yearcite{n159_fukui}; \cite{n159_tokuda}).

The geometry in figure \ref{fig:boundary_res}b suggests that gas collision occurred in the north region prior to the south region.
In order to verify such a difference in the timing of the gas collision, we compare the evolutionary stages of giant molecular clouds (GMCs) with the crossing timescale of the gas collision.
\citet{kawamura} classified GMCs in the LMC into four evolutionary stages, Type $\rm{\,I\,}$--$\rm{I\hspace{-.15em}I\hspace{-.15em}I}$ and last stage (Type $\rm{\,I\,}$ is the youngest phase).
They found the evolutionary sequence from the south (young) to the north (old) of the  H\,\emissiontype{I} ridge region, i.e., Type-$\rm{\,I\,}$ GMCs in region 3, Type-$\rm{I\hspace{-.15em}I\hspace{-.15em}I}$ GMCs around N159 and last-stage GMCs near 30 Dor.
The transition timescale from Type $\rm{\,I\,}$ to last stage is $\sim 25$ Myr.
Assuming that the velocity of the inflow gas (L-component) is 100 km/s and the thickness of the LMC disk is 2 kpc (\cite{thickness}), it spends 20 Myr for the inflow gas to cross the LMC disk, which is consistent with the transition timescale ($\sim 25$ Myr) from Type-$\rm{\,I\,}$ GMCs in region 3 to last-stage GMCs near 30 Dor.
Therefore, the colliding cloud geometry shown in figure \ref{fig:boundary_res}b can reasonably explain the evolutionary sequence in the H\,\emissiontype{I} ridge region found by \citet{kawamura}.
%
%
%
\subsection{Comparison of dust extinction with dust emission}
In order to check the validity of the geometry that the L-component in region 3 is located behind the LMC disk, we compare the dust extinction with the dust emission; the dust extinction is caused by only the dust in front of the stars while the emission can come from all dust.
As a dust emission map, we use the map of the dust optical depth at 353 GHz, $\tau_{353}$, obtained from the Planck and IRAS far-infrared data (\cite{tau_map}).
To estimate $\tau_{353}$ in the LMC, \citet{tsuge} calculated the Galactic foreground dust optical depth from the foreground  H\,\emissiontype{I} intensity ($W$(H\,\emissiontype{I})) using the conversion factor from $W$(H\,\emissiontype{I}) to $\tau_{353}$ derived from \citet{tau_h1}.
We use the $\tau_{353}$ map after subtraction of the foreground component.

\begin{figure*}
 \begin{center}
  \includegraphics[width=16cm]{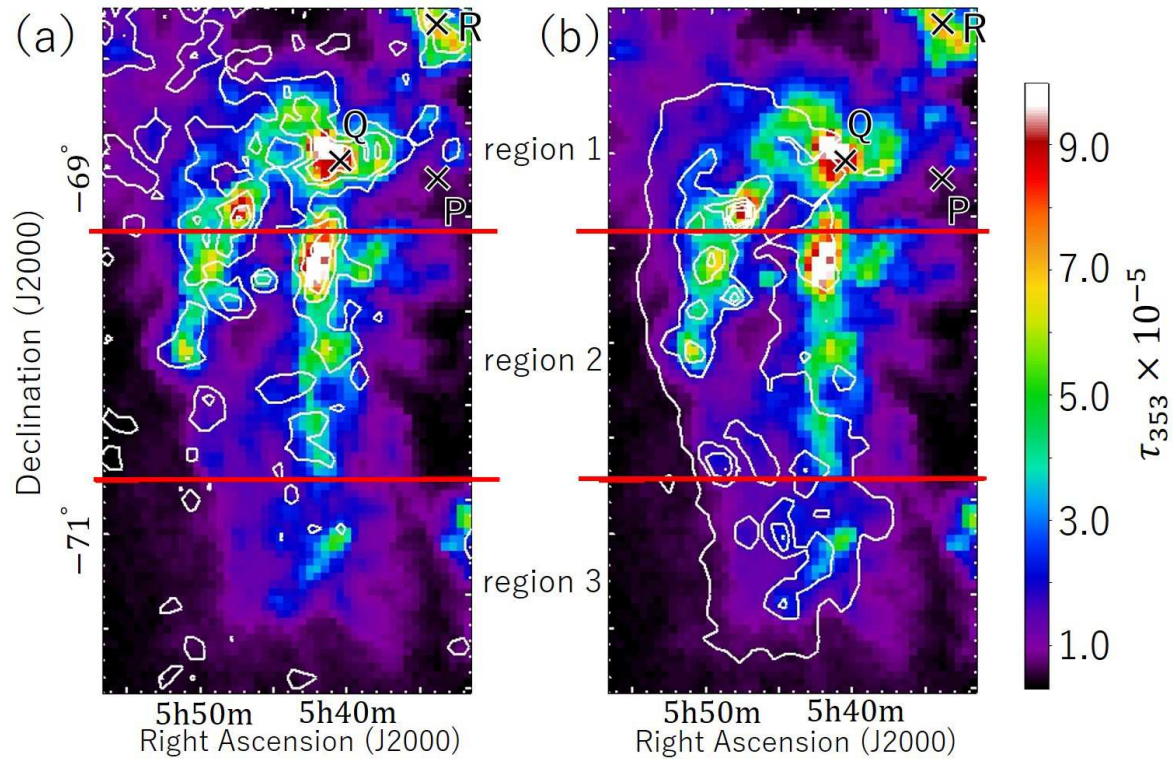} 
 \end{center}
\caption{Dust optical depth at 353 GHz, $\tau_{353}$ map superposed on the contours of (a) $A_V$(80\%) detected with significance higher than $1\sigma$ and $A_V \geq$0.6 mag, and (b) $N$(H) of the L-component which are the same contours as in figure \ref{fig:av_w_gas}a. The contour levels in panel (a) are 0.6, 1.0, 1.3 and 1.7 mag. The red horizontal lines represent boundaries 1 and 2. The cross symbols are the same as those in figure \ref{fig:av_w_gas}.}\label{fig:tau_w_av}
\end{figure*}

Figure \ref{fig:tau_w_av}a shows the $\tau_{353}$ map with the contours of $A_V$(80\%) detected with significance higher than 1$\sigma$ and $A_V \geq 0.6$ mag, from which we find that the dust emission exists in region 3 while the dust extinction does not.
In regions 1 and 2, the dust emission correlates well with the dust extinction.
On the other hand, figure \ref{fig:tau_w_av}b shows the $\tau_{353}$ map with the contours of $N$(H) of the L-component, from which we find that the dust emission correlates well with the L-component gas.
$N$(H) of the L-component traces the dust emission in region 3.
Those results strongly support that the L-component in region 3 is located behind the LMC.
%
%

\begin{figure}
 \begin{center}
  \includegraphics[width=8cm]{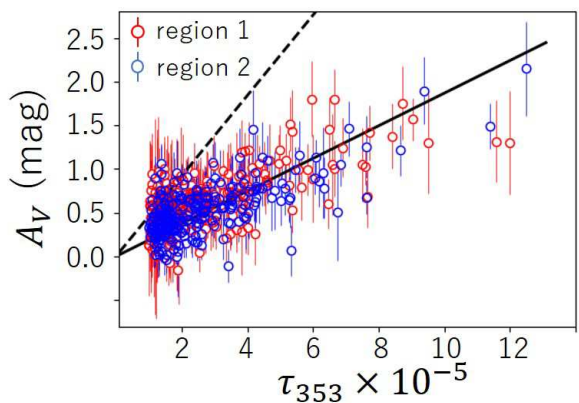} 
 \end{center}
\caption{Scatter plot between $A_V$ and $\tau_{353}$ for regions 1 (red) and 2 (blue). Black solid line represents the result of the least-square fit to all the data points, while black dotted line represents the relation between $A_V$ and $\tau_{353}$ for the diffuse Galactic dust.}\label{fig:av_vs_tau}
\end{figure}

Now that the dust clouds in regions 1 and 2 are likely located in front of almost all the stars, we can securely investigate the relation between the dust extinction and emission in those regions.
In figure \ref{fig:av_vs_tau}, we present the scatter plot of $A_V$ and $\tau_{353}$ for each pixel of regions 1 and 2 with $\tau_{353}> 1.0 \times 10^{-5}$ for both regions.
The figure shows that there is no systematic difference in the $A_V$-$\tau_{353}$ relation between regions 1 and 2.
The black solid line is the result of the least-square fit to all the data points in regions 1 and 2, assuming that the relation between $A_V$ and $\tau_{353}$ has a zero-point intercept.
The estimated slope is $A_V$/$\tau_{353}=(1.86 \pm 0.04)\times 10^{4}$.
\citet{tau_map} gives $E(B-V)/\tau_{353}=1.5\times 10^4$ for the Galactic diffuse dust ($\tau_{353}<5\times10^{-6}$).
Assuming $R_V=3.1$, we have $A_V/\tau_{353}=4.6\times 10^4$ for the Galactic diffuse dust, which is 2.4 times larger than our estimation of $A_V$/$\tau_{353}=(1.86 \pm 0.04)\times 10^{4}$ for the dust in regions 1 and 2 of the LMC H\,\emissiontype{I} ridge region.
The same trend was reported for our Galaxy by \citet{tau_map}, who found that $\tau_{353}/N$(H) systematically increases by a factor of $\sim 2$ from the diffuse to denser regions due to dust aggregation to larger grains.
Therefore, the relation between $A_V$ and $\tau_{353}$ estimated in this paper is consistent with that in the dense ISM in our Galaxy, which suggests that the properties of large grains in the clouds of the I- and L-components are similar to those in our Galaxy. 
%
%

\subsection{Dust/gas ratio and CO-to-$\rm H_2$ conversion factor}
The $A_V$/$N(\rm H)$ values at $X=40\%$ of the L-, I- and D-components are 1/10, 1/4 and 1/5, respectively, of the Galactic $A_V$/$N(\rm H)$ values of $\sim 5.3\times10^{-22}$ mag/(H $\rm cm^{-2}$), while those at $X=80\%$ are 1/5, 1/4 and 1/3 of the Galactic values, respectively.
$A_V$/$N(\rm H)$ at $X=40\%$ of all the components is consistent with that derived by \citet{furuta} using the averaged dust extinction map (see table \ref{tab:fit_res} and figure \ref{fig:fit_res}a in the present paper).
On the other hand, $A_V$/$N(\rm H)$ at $X=80\%$ of the L-component is 2 times higher than that derived by \citet{furuta}.
This result can be interpreted as follows: since the L-component is penetrating the LMC disk (figure \ref{fig:boundary_res}b), the averaged dust extinction of \citet{furuta} would not trace all the dust in the L-component existing along the line of sight, which leads to under-estimation of the dust extinction.
 $A_V$/$N(\rm H)$ at $X=80\%$ of the D-component is also 2 times higher than that derived by \citet{furuta}.
 This reasonably indicates that stars are randomly distributed between the foreground and background, and thus the averaged dust extinction would correspond to $\sim$1/2 of the total dust extinction.

Based on the scenario that the low metalliciy gas from the SMC is accreted to and mixed in the H\,\emissiontype{I} ridge region (\cite{fukui_2017}; \cite{furuta}), we compare the resultant $A_V$/$N(\rm H)$ at $X=80\%$ of each velocity component with the gas-to-dust ratios (GDRs) of the LMC and the SMC.
Using Spitzer far-infrared, ACTA+Parkes H\,\emissiontype{I} 21 cm and NANTEN CO data, \citet{gdr_lmc_spitzer} and \citet{gdr_smc_spitzer} obtained the averaged GDR about 3 times higher for the LMC and $\sim$10 times higher for the SMC, respectively, as compared to the Galactic GDR.
\citet{gdr_planck} calculated the dust optical depth from the Planck data and suggested that the GDRs of the LMC and the SMC are 2.4 times and 13 times higher than the Galactic GDR, respectively.
\citet{gdr_herschel} evaluated the GDRs in the Magellanic Clouds from dust and gas surface densities based on Herschel far-infrared, H\,\emissiontype{I} 21 cm, CO and $\rm{H{\alpha}}$ observations, and obtained the GDRs 3 times higher for the LMC and 6 times higher for the SMC than the Galactic value.
In summary, the previous studies based on the dust emission indicate that the GDR is 2--3 times higher for the LMC and 6--13 times higher for the SMC than the Galactic value.

The $A_V$/$N(\rm H)$ value at $X$=80\% estimated for the D-component in the present study corresponds to $\sim$1/3 of the Galactic value (see figure \ref{fig:fit_res}a), which is consistent with the GDR of the LMC as mentioned above.
This implies that our method works well to evaluate the dust extinction caused by the D-component gas.
The $A_V$/$N(\rm H)$ values estimated for the L- and I-components in the present study correspond to $\sim$1/5 of the Galactic value (figure \ref{fig:fit_res}a), which are intermediate values between the GDRs of the SMC and the LMC as mentioned above.
Thus we suggest that the L- and I-component gases are mixed with the D-component gas due to the interaction between the LMC gas and the inflow gas from the SMC.
Assuming that $A_V$/$N(\rm H)$ of the inflow gas was originally 1/10 of the Galactic value, 1/5 of the Galactic $A_V$/$N(\rm H)$ for the L- and I-components can be explained by a mixture of the inflow gas with the LMC gas having 1/3 of the Galactic $A_V$/$N(\rm H)$ at a mass ratio of 1:1.
This result is consistent with the mass fraction of the inflow gas ($\sim$50\%) in the H\,\emissiontype{I} ridge region expected from the ratio of $\tau_{353}$ to the H\,\emissiontype{I} intensity (\cite{fukui_2017}).

As can be seen in figure \ref{fig:fit_res}b, the $X_{\rm CO}$ factors of each velocity component stay constant within the errors over the whole range of $X$\% and are consistent with those in \citet{furuta} (see table \ref{tab:fit_res}).
This result indicates that the abundance ratio of CO to $\rm H_2$ is relatively uniform within the clouds associated with each velocity component.
The $X_{\rm CO}$ factors of the I- and D-components are similar to that of the LMC $\sim 2.9\times10^{20}$ ${\rm cm^{-2}}/(\rm{K\ km\ s^{-1}})$ obtained by \citet{corr_xco}, who correct the effect of the photodissociation of CO due to the lack of dust shielding.
In addition, the $X_{\rm CO}$ factor of the L-component is in the range from that of the LMC $\sim 2.9\times10^{20}$ ${\rm cm^{-2}}/(\rm{K\ km\ s^{-1}})$ to that of the SMC $\sim 7.6\times10^{20}$ ${\rm cm^{-2}}/(\rm{K\ km\ s^{-1}})$ obtained by \citet{corr_xco}, which also supports the above picture of the inflow gas mixed with the LMC.

As a whole, the dust geometry and the difference in $A_V$/$N$(H) and $X_{\rm CO}$ between the three velocity components suggest the galactic interaction between the LMC and the SMC in the H\,\emissiontype{I} ridge region.
In our future work, we will apply this method to the entire LMC field to discuss a possibility of the overall massive star formation triggered by the galactic interaction.
\section{Conclusion}
We develop the method to evaluate $A_V$ used in \citet{furuta} by taking percentile values based on the $X$ percentile method proposed by \citet{dobashi}.
Using our new method, we construct the three-dimensional $A_V$ map of the LMC H\,\emissiontype{I} ridge region. 
From comparison of $A_V$ along the line of sight with $N$(H) of different velocities, we decompose $A_V$ into three velocity components, and evaluate the three-dimensional dust geometry and dust/gas ratios of the three velocities.
Our main results are as follows:
\begin{enumerate}
\item The resultant dust geometry suggests that the L-component in the north region is penetrating the LMC disk, while that in the south region is located behind the LMC disk.
In addition, the I-component (i.e., intermediate velocity) is located in front of almost all the stars in the LMC.
This can be explained by the deceleration of the L-component through interaction with the D-component.
These results indicate that the gas collision is on-going in the north prior to the south of the H\,\emissiontype{I} ridge region.
Such a difference in the timing of the gas collision is consistent with the evolutionary sequence of molecular clouds in the H\,\emissiontype{I} ridge region found by \citet{kawamura}.

\item We compare the spatial distributions of the dust extinction $A_V(80\%)$ and the dust emission $\tau_{353}$ in the H\,\emissiontype{I} ridge region.
We find that they are correlated well in the north while they are not in the south.
The dust extinction is not detected whereas the dust emission exists and is well traced by $N$(H) of the L-component in the south region.
Those results strongly support the geometry that the L-component in the south region is located behind the LMC disk.
The relation between $A_V$ and $\tau_{353}$ suggests the dominance of large dust grains in the clouds of the I- and L-components, as found in dense clouds in our Galaxy.

\item $A_V$(80\%)/$N$(H) of the D-component (Disk, see figure \ref{fig:boundary_res}b) is $\sim$1/3 of the Galactic value, which is consistent with the gas-to-dust ratio (GDR) of the LMC estimated from the dust emission.
On the other hand, $A_V$(80\%)/$N$(H) of the L- and I-components is $\sim$1/5 of the Galactic value, which is an intermediate GDRs between the SMC and the LMC.
This result suggests that the inflow gas from the SMC is mixed with the LMC gas at a mass ratio of 1:1 through the interaction between Magellanic Clouds in the H\,\emissiontype{I} ridge region.
The difference in the $X_{\rm CO}$ factors of each velocity component suggests a result consistent with the above picture.
\end{enumerate}
As a whole, our results support the scenario that massive star formation in 30 Dor has been triggered by the tidal interaction between the LMC and the SMC.

\begin{ack}
The IRSF project is a collaboration between Nagoya University and the SAAO supported by the Grants-in-Aid for Scientific Research on Priority Areas (A) (Nos. 10147207 and 10147214) and Optical \& Near-Infrared Astronomy Inter-University Cooperation Program, from the Ministry of Education, Culture, Sports, Science and Technology (MEXT) of Japan and the National Research Foundation (NRF) of South Africa.
This research was financially supported by Grant-in-Aid for JSPS Fellows Grant Number 20J12119.
\end{ack}



\begin{thebibliography}{99}
\bibitem[Balbinot et al.(2015)]{thickness}
Balbinot, E., et al.\ 2015, \mnras, 449, 1129. doi:10.1093/mnras/stv356
\bibitem[Bekki \& Chiba(2007a)]{bekkia} 
Bekki, K. \& Chiba, M.\ 2007a, PASA, 24, 21. doi:10.1071/AS06023
\bibitem[Bekki \& Chiba(2007b)]{shock_sim} 
Bekki, K., \& Chiba, M.\ 2007b, \mnras, 381, L16
\bibitem[Bernard et al.(2008)]{gdr_lmc_spitzer} 
Bernard, J.-P., et al.\ 2008, \aj, 136, 919
\bibitem[Bessell \& Brett(1988)]{tr_giant} 
Bessell, M.~S., \& Brett, J.~M.\ 1988, \pasp, 100, 1134
\bibitem[Bohlin et al.(1978)]{galactic_dustgas}
Bohlin R.~C., Savage B.~D., \& Drake J.~F., 1978, ApJ, 224, 132
\bibitem[Cardelli, Clayton and Mathis(1989)]{red_law2}
Cardelli, J.~A., Clayton, G.~C., \& Mathis, J.~S.\ 1989, \apj, 345, 245
\bibitem[Crowther et al.(2010)]{30dor_star}
Crowther, P.~A., Schnurr, O., Hirschi, R., Yusof N., Parker R.~J., Goodwin S.~P., \& Kassim H.~A.\ 2010, \mnras, 408, 731
\bibitem[Crowther et al.(2016)]{crowther_2016}
Crowther, P.~A., et al.\ 2016, \mnras, 458, 624. doi:10.1093/mnras/stw273
\bibitem[De Marchi \& Panagia(2014)]{30dor}
De Marchi, G., \& Panagia, N.\ 2014, \mnras, 445, 93
\bibitem[Dickey \& Lockman(1990)]{nh_conv}
Dickey, J.~M., \& Lockman, F.~J.\ 1990, \araa, 28, 215
\bibitem[Dobashi et al.(2008)]{dobashi}
Dobashi, K., Bernard, J.-P., Hughes, A., Paradis, D., Reach, W. T., \& Kawamura, A.\ 2008, \aap, 484, 205
\bibitem[Doran et al.(2013)]{r136_star}
Doran, E.~I., Crowther, P.~A., de Koter, A., et al.\ 2013, \aap, 558, A134
\bibitem[Elias et al.(1985)]{color_corr2} 
Elias, J.~H., Frogel, J.~A., \& Humphreys, R.~M.\ 1985, \apjs, 57, 91
\bibitem[Fukui et al.(1999)]{comap} 
Fukui, Y., et al.\ 1999, \pasj, 51, 745
\bibitem[Fukui et al.(2008)]{h2_conv} 
Fukui, Y., et al.\ 2008, \apjs, 178, 56
\bibitem[Fukui et al.(2015)]{tau_h1}
Fukui Y., Torii K., Onishi T., Yamamoto H., Okamoto R., Hayakawa T., Tachihara K., \& Sano H.,\ 2015, \apj, 798, 6
\bibitem[Fukui et al.(2017)]{fukui_2017} 
Fukui, Y., Tsuge, K., Sano, H., Bekki, K., Yozin, C., Tachihara, K., \& Inoue, T.\ 2017, \pasj, 69, L5
\bibitem[Fukui et al.(2019)]{n159_fukui} 
Fukui Y., et al., 2019, ApJ, 886, 14
\bibitem[Furuta et al.(2019)]{furuta}
Furuta, T., Kaneda, H., Kokusho, T.,  Ishihara D., Nakajima Y., Fukui Y., \& Tsuge K.\ 2019, \pasj, 71, 95
\bibitem[Indebetouw et al.(2020)]{30dor_alma}
Indebetouw R., Wong T., Chen C.-H.~R., Kepley A., Lebouteiller V., Madden S., \& Oliveira J.~M., 2020, ApJ, 888, 56
\bibitem[Imara \& Blitz(2007)]{extinction_nice} 
Imara, N., \& Blitz, L.\ 2007, \apj, 662, 969
\bibitem[Kato et al.(2007)]{irsf}
Kato, D., et al.\ 2007, \pasj, 59, 615
\bibitem[Kaluzny et al.(1998)]{variable_1}
Kaluzny J., Stanek K.~Z., Krockenberger M., Sasselov D.~D., Tonry J.~L., \& Mateo M. \ 1998, \aj, 115, 1016
\bibitem[Kawamura et al.(2009)]{kawamura}
Kawamura, A., et al.\ 2009, \apjs, 184, 1. doi:10.1088/0067-0049/184/1/1
\bibitem[Kim et al.(2003)]{himap}
Kim, S., Staveley-Smith, L., Dopita, M.~A., Sault, R.~J., Freeman, K.~C., Lee, Y. \& Chu, Y.\ 2003, \apjs, 148, 473
\bibitem[Lada et al.(1994)]{nice}
Lada, C.~J., Lada, E.~A., Clemens, D.~P., \& Bally, J.\ 1994, \apj, 429, 694
\bibitem[Leroy et al.(2007)]{gdr_smc_spitzer}
Leroy A., Bolatto A., Stanimirovic S., Mizuno N., Israel F., \& Bot C.,\ 2007, \apj, 658, 1027
\bibitem[Lombardi \& Alves(2001)]{nicer}
Lombardi, M., \& Alves, J.\ 2001, \aap, 377, 1023
\bibitem[Nakajima et al.(2005)]{color_corr}
Nakajima, Y., et al.\ 2005, \aj, 129, 776
\bibitem[Meixner et al.(2006)]{mips} 
Meixner, M., et al.\ 2006, \aj, 132, 2268
\bibitem[Pietrzy{\'n}ski et al.(2019)]{proximity}
Pietrzy{\'n}ski, G., Graczyk, D., Gallenne, A., et al.\ 2019, \nat, 567, 200
\bibitem[Pineda et al.(2017)]{corr_xco}
Pineda, J.~L., et al.\ 2017, \apj, 839, 107
\bibitem[Planck collaboration(2011)]{gdr_planck} 
Planck Collaboration,\ 2011, \aap, 536, A17
\bibitem[Planck Collaboration(2014)]{tau_map}
Planck Collaboration,\ 2014, \aap, 571, A11
\bibitem[Roman-Duval et al.(2014)]{gdr_herschel}
Roman-Duval, J., et al.\ 2014, \apj, 797, 86
\bibitem[Schneider et al.(2018)]{schneider} 
Schneider, F.~R.~N., et al.\ 2018, Science, 359, 69. doi:10.1126/science.aan0106
\bibitem[Tatton et al.(2013)]{tatton} 
Tatton, B.~L., et al.\ 2013, \aap, 554, A33
\bibitem[Tokuda et al.(2019)]{n159_tokuda}
Tokuda K., et al., 2019, ApJ, 886, 15
\bibitem[Tsuge et al.(2019)]{tsuge}
Tsuge, K., et al.\ 2019, \apj, 871, 44
\bibitem[Tsuge et al.(2021)]{tsuge_2021} 
Tsuge K., et al., 2021, \apj, submitted (arXiv:2010.08816)
\bibitem[van der Marel \& Cioni(2001)]{inclination}
van der Marel, R.~P., \& Cioni, M.-R.~L.\ 2001, \aj, 122, 1807
\bibitem[Westerlund(1997)]{metal}
Westerlund, B.~E.\ 1997, The Magellanic Clouds (New York: Cambridge Univ. Press), 243
\bibitem[Yao et al.(2015)]{variable_2}
Yao, X., et al.\ 2015, \aj, 150, 107
\bibitem[Yozin \& Bekki(2014)]{yozin}
Yozin C., Bekki K., 2014, MNRAS, 443, 522. doi:10.1093/mnras/stu1132







\end{thebibliography}
\end{document}